\documentclass[%
 reprint,
 amsmath,amssymb,
 aps,
 prx,
]{revtex4-2}

\usepackage{graphicx}
\usepackage{amsfonts}
\usepackage{dcolumn}
\usepackage{bm}
\usepackage{lipsum}
\usepackage[colorlinks=true,citecolor=blue,linkcolor=blue,urlcolor=magenta]{hyperref}
\usepackage{bbold}
\usepackage{mathtools}
\usepackage{longtable}
\usepackage{tabularx}
\usepackage{svg}
\hypersetup{breaklinks=true}

\newcommand{\aspeed}[2]{\nu_{#1,#2}}

\DeclareMathOperator\erf{erf}

\DeclareMathOperator\Cov{Cov}
\DeclareMathOperator\Evalue{\mathbb{E}}
\DeclareMathOperator\Var{Var}
\DeclareMathOperator*{\argmax}{arg\,max}

\renewcommand{\Re}{\operatorname{\mathbb{R}e}}
\renewcommand{\Im}{\operatorname{\mathbb{I}m}}

\begin{document}

\preprint{APS/123-QED}

\title{Laboratory Constraints on the Neutron-Spin Coupling of feV-scale Axions}

\author{Junyi Lee}
\affiliation{Institute of Materials Research and Engineering, Agency for Science, Technology and Research (A*STAR), 2 Fusionopolis Way, \#08-03 Innovis, Singapore 138634, Singapore}

\author{Mariangela Lisanti}
\affiliation{Department of Physics, Princeton University, Princeton, NJ 08544, USA}
\affiliation{Center for Computational Astrophysics, Flatiron Institute, New York, NY 10010, USA}

\author{William A. Terrano}
\affiliation{Department of Physics, Arizona State University, Tempe, AZ 85287, USA}

\author{Michael Romalis}
\affiliation{Department of Physics, Princeton University, Princeton, NJ 08544, USA}

\date{\today}

\begin{abstract}
Ultralight axion-like particles can contribute to the dark matter near the Sun, leading to a distinct, stochastic signature in terrestrial experiments. We search for such particles through their neutron-spin coupling by re-analyzing approximately 40 days of data from a K-$^3$He co-magnetometer with a new frequency-domain likelihood-based formalism that properly accounts for stochastic effects over all axion coherence times relative to the experimental time span. Assuming that axions make up all of the dark matter in the Sun's vicinity, we find a median 95\% upper limit on the neutron-spin coupling of $2.4 \times 10^{-10}$~GeV$^{{-1}}$ for most axion masses from 0.4 to 4 feV, which is about five orders of magnitude more stringent than previous laboratory bounds in that mass range. Although several peaks in the experiment's magnetic power spectrum suggest the rejection of a white-noise null hypothesis, further analysis of their lineshapes yields no positive evidence for a dark matter axion.
\end{abstract}

\maketitle

\section{Introduction\label{sec:intro}}

Although the existence of dark matter is supported by astrophysical and cosmological observations over many length scales~\cite{Corbelli2000, Clowe2006, Jee2007, Jarosik2011}, its fundamental nature remains a mystery. Ultra-light pseudo-scalar particles that are the pseudo-Nambu-Goldstone bosons of a spontaneously broken global symmetry arise generically in many theories beyond the Standard Model~\cite{Wise1981, Nilles1982, Svrcek2006, Arvanitaki2010} and can potentially contribute to the local dark matter. These ``axions'' can couple to the fermionic sector and lead to a neutron-spin coupling. In this paper, we reanalyze data from a Princeton-based K-$^3$He co-magnetometer experiment~\cite{Vasilakis2009} to obtain leading laboratory constraints on the neutron-spin coupling of feV-scale axions. 

One well-motivated example of an ultra-light pseudo-scalar particle is the QCD axion. It was originally proposed to solve the strong CP problem within QCD~\cite{Peccei1977_PRL, Peccei1977_PRD, Weinberg1978, Wilczek1978}, but is also an attractive dark matter candidate~\cite{Dine1983, Abbott1983, Preskill1983, Sikivie2021, Kwon2021, Bartram2021}. While the particle mass and coupling of the QCD axion are fundamentally related, this is not the case for more generic axions where the coupling strength and mass are independent of each other.  Indeed, axions generically couple to neutrons with the Lagrangian
\begin{equation}
    \mathcal{L}_{\rm int} = g_\text{aNN} \bar{N}\gamma^\mu \gamma_5 N\, \partial_\mu a \,,
    \label{eq:L_int}
\end{equation}
where $g_\text{aNN}$ parametrizes the coupling strength and $a, N$ refer to the axion and neutron, respectively. 
The interaction in Eq.~\eqref{eq:L_int} mediates a  spin-dependent long-range force, which has been searched for by~\cite{Hoyle:2004cw, Kapner:2006si, Adelberger2007, Glenday2008, Vasilakis2009}. However, this signal is suppressed by two factors of the (small) coupling constant $g_\text{aNN}$.

An alternate detection strategy is to search for dark matter axions that couple to the nuclear spin in a detector target~\cite{Abel2017, Teng2019, Garcon2019, Jiang2021, Bloch2022}. In this case, the interaction Hamiltonian, which follows from the non-relativistic limit of Eq.~\eqref{eq:L_int}, is only suppressed by one factor of $g_\text{aNN}$:
\begin{equation}
    H_{\rm int} = g_\text{aNN} \boldsymbol{\nabla}a \cdot \boldsymbol{\sigma}_N \,,
    \label{eq:H_int}
\end{equation}
where $\boldsymbol{\sigma}_N$ is the neutron spin. Because light dark matter axions near the Sun are sufficiently numerous to be treated as an oscillating classical field, interference effects between different axion waves can lead to stochastic signals in terrestrial experiments~\cite{Foster2018, Derevianko2018,Lisanti2021}, which must be accounted for to properly recover or constrain a signal. This effect was first discussed by~\cite{Foster2018, Derevianko2018} for the axion-photon coupling that many haloscope experiments are based on \cite{Salemi2021, Braine2020, Zhong2018, Gramolin2021, McAllister2017}, but was only recently addressed for the axion-nuclear spin coupling~\cite{Lisanti2021}.  For the latter case, the signal is proportional to the gradient of the axion field, and it also depends on the direction of $\boldsymbol{\sigma}_N$, which for a terrestrial experiment will oscillate with a period of a sidereal day as the Earth rotates about its axis.

Experimentally, the interaction in Eq.~\eqref{eq:H_int} should lead to a measurable spin-dependent energy shift in the valence neutron of $^3$He that rotates the macroscopic magnetization of spin-polarized $^3$He atoms. In a K-$^3$He co-magnetometer, this spin polarization is created along a circularly polarized pump beam and the magnetization of the atoms is probed by a separate orthogonal probe beam. In this paper, we report novel laboratory constraints of the axion's neutron-spin coupling based on analyzing approximately 40 days of raw magnetization orientation data in a K-$^3$He co-magnetometer that was originally used in an exotic long-range force experiment \cite{Vasilakis2009}.

Our analysis fully accounts for the stochastic nature of dark matter axions using a new frequency-domain likelihood-based formalism, which is complementary to the time-binned methodology introduced in the theoretical counterpart to this work~\cite{Lisanti2021}. The likelihood formalism in \cite{Lisanti2021} properly accounts for all two-point correlations of the axion field in the time-domain data. As explicitly demonstrated in that work, ignoring these correlations can result in failure to discover a signal and/or result in unreliable upper limits. We explicitly show in Appendix~\ref{sec:validation} that both the frequency and time-binned approaches yield similar results when applied on the same subset of simulated and experimental data. The analysis in this paper, like that in \cite{Lisanti2021}, works for all axion coherence times independent of the experimental time span (including in the regime when the experimental measurement time is shorter or comparable to the axion coherence time), which is a feature not achievable in previous frameworks~\cite{Centers2020}. Crucially, given a measured experimental spectrum, our formalism sets limits on (or recovers, in the case of a discovery) $g_\textrm{aNN}$ in a way that fully takes into account the stochasticity of the dark matter axion signal. This is in contrast with a recent work that studied the stochastic axion lineshape without providing a methodology for correctly extracting $g_\textrm{aNN}$ from an experimental measurement \cite{Gramolin2022}. We have rigorously validated our analysis by applying it to a Monte-Carlo ensemble of stochastic axion signals to ensure that it indeed recovers or sets the correct limits on $g_\textrm{aNN}$ (see Appendix~\ref{sec:validation}). There is therefore no need to apply any of the corrections in \cite{Centers2020} to our limits.

This paper should be contrasted with a recent analysis~\cite{Bloch2020, Bloch2022} that used the published power spectra of the Princeton-based co-magnetometer (not the raw underlying data) without properly treating the signal's stochasticity, compromising its ability to reliably recover or set upper limits. Our analysis of the raw data also allows us to examine and subsequently reject peaks in the spectrum that may be plausible axion candidates.

Assuming that the local dark matter density is wholly due to axions, our median 95\% upper limit on the axion's neutron-spin coupling is $g_\text{aNN} < 2.4 \times 10^{-10}$~GeV$^{-1}$ for axion masses from 0.4 to 4~feV. This is about five orders of magnitude stronger than previous laboratory bounds in the same mass range.  For example, the original long-range force experiment in~\cite{Vasilakis2009} reported a 95\% upper limit of $g_\text{aNN} < 9.1 \times 10^{-5}$~GeV$^{-1}$ and a recent terrestrial experiment searching for dark matter axions in a similar mass range reported a tightest 90\% upper limit of $4.1 \times 10^{-5}$~GeV$^{-1}$~\cite{Garcon2019}.  Close to the largest axion mass we analyzed (41 feV), another experiment recently reported a more comparable 95\% upper limit of $3.2 \times 10^{-9}$~GeV$^{-1}$ at 53 feV~\cite{Jiang2021}, although we note that for the quoted limit the analysis there has significant frequency gaps (compared to the axion's linewidth) and also does not account for the axion's stochasticity. At the largest mass we analyzed, the NASDUCK collaboration recently reported a 95\% upper limit of $10^{-6}$~GeV$^{-1}$ \cite{Bloch2022}, which is about two orders of magnitude weaker than our results. Our constraints are comparable with limits derived from stellar cooling arguments although the latter are subject to significant astrophysical uncertainties. Our results therefore serve as useful complementary bounds.

The magnetic power spectrum of the original experiment in~\cite{Vasilakis2009} was only optimized for a narrow bandwidth about 0.17 Hz and peaks outside that narrow bandwidth were not seriously investigated. Our likelihood analysis of those peaks, which tests an axion signal hypothesis against a null white background hypothesis, yields a few peaks with significance above 5$\sigma$ (after taking into account the look-elsewhere effect) that persist over the entire data-taking duration. Although these results suggest that the white background model should be rejected for these peaks, accepting an axion signal hypothesis should require additional tests. Comparison with the expected lineshapes from Monte-Carlo simulations suggest a non-axion origin for all these peaks.

This paper is structured as follows. In Sec.~\ref{sec:freq_domain}, we review features of the frequency-domain likelihood analysis, leaving detailed derivations to Appendix~\ref{sec:derivation}. We then describe the original experiment of~\cite{Vasilakis2009} and its pertinent features to this work in Sec.~\ref{sec:exp_setup}, before discussing the results of the analysis in Secs.~\ref{sec:results}--\ref{sec:peaks} and concluding in Sec.~\ref{sec:conclusion}. We also include two additional appendices. In Appendix~\ref{sec:validation}, we present the validation of our frequency-domain likelihood analysis with the method of \cite{Lisanti2021}, and in Appendix~\ref{sec:freq_grid}, we motivate the frequency resolution at which we test for axion candidates.

\section{Frequency-domain analysis\label{sec:freq_domain}}

\subsection{Axion Signal Modeling}

Weakly interacting astrophysical axions confined within a volume $V$ can be treated as a free scalar field
\begin{equation}
a(x) = \sum_\mathbf{p} \frac{1}{\sqrt{2 V \omega_\mathbf{p}}} \left(b_\mathbf{p} \, e^{-ip\cdot x} + b_\mathbf{p}^\dagger \, e^{ip\cdot x} \right),
\end{equation}
where the sum is over all three-momenta $\mathbf{p}$, and $b_\mathbf{p}, b_\mathbf{p}^\dagger$ are the annihilation and creation operators of the mode with energy $\omega_\mathbf{p}$ and four-momentum $p$. If the local dark matter density is due wholly to axions, then their occupancy numbers must be extremely large and we may therefore take the classical limit where $b^\dagger_\mathbf{p} b_\mathbf{p}= N_\mathbf{p}$ is the mean occupation number of the mode with momentum~$\mathbf{p}$. Doing so, we obtain the classical axion field
\begin{equation}
a(x) = \sum_\mathbf{p} \sqrt{\frac{2 N_\mathbf{p}}{V \omega_\mathbf{p}}} \cos(p^0 t - \mathbf{p}\cdot\mathbf{x} + \phi_\mathbf{p}) \,,
\label{eq:classical_axion_field}
\end{equation}
where $\phi_\mathbf{p}$ is an unknown phase between 0 and 2$\pi$ that can be modeled as uniformly distributed over that interval. The sum over all modes $\mathbf{p}$ with distinct energies $\omega_\mathbf{p}$ leads to interference effects noticeable after a characteristic coherence time $\tau_a$, which is dependent on the momentum distribution of the axions. For example, in the Standard Halo Model, axions are assumed to follow a Maxwell-Boltzmann velocity distribution in the Galactic rest frame,
\begin{equation}
f(\mathbf{v})\,d^3v =
\dfrac{1}{(2\pi\sigma_\textrm{v}^2)^{3/2}}\exp\left(-\dfrac{(\mathbf{v} + \mathbf{v}_E)^2}{2\sigma_\textrm{v}^2}\right)\, d^3v \,,
\label{eq:mb_velocity_dist}
\end{equation}
where $\mathbf{v}$ is the axion velocity in the laboratory frame, $\mathbf{v}_E$ is the Earth's velocity \footnote{In our analysis, we take the average Earth velocity over the time period of the experiment, including Earth's orbital velocity about the Sun, but ignoring Earth's rotation about its axis. This was shown to be a minor correction in \cite{Lisanti2021}.} so that $\mathbf{v} + \mathbf{v}_E$ is the axion velocity in the Galactic rest frame, and $\sqrt{2}\,\sigma_\textrm{v} \sim 220$~km/s is the distribution's modal speed, which is typically taken to be the Sun's circular velocity in the Galactic rest frame~\cite{Catena2012, Bovy2012, Eilers2019}. In this case, the characteristic coherence time for an axion with mass $m_a$ may be defined as $\tau_a=1/(\sigma_\textrm{v}^2 m_a)$~\cite{Derevianko2018}.

\begin{figure}
    \centering
    \includegraphics[width=0.49\textwidth]{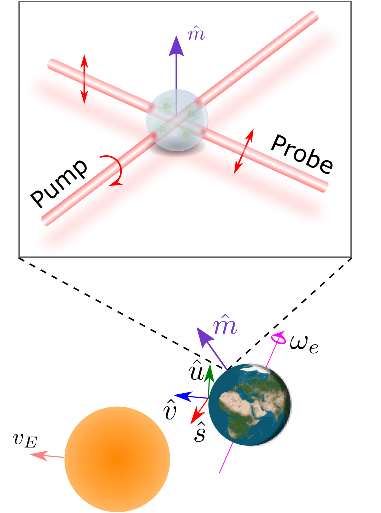}
    \caption{(color online) Schematic showing the sensitive axis $\hat{\mathbf{m}}$ of the experiment as an outward normal from the surface of the Earth that rotates at the sidereal frequency $\omega_e$, leading to sidebands of the axion's power spectrum. The inset shows a schematic of the co-magnetometer, which consists of spin-polarized atoms in a glass vapor cell that was optically pumped by resonant circularly polarized light and probed by off-resonant linearly polarized light in the horizontal plane. As described in the main text, this implies that the sensitive axis $\hat{\mathbf{m}}$ is oriented vertically and by convention, we take it to be the outward normal from the surface of the Earth. It is convenient, when deriving the covariance matrix of the axion's frequency spectrum, to use an orthonormal coordinate system $\{\hat{s}, \hat{u}, \hat{v}\}$ such that $\hat{v}$ is parallel to the Earth's velocity $v_E$ (see Appendix \ref{sec:deriv_integration}), which is dominated by the Sun's circular velocity.}
    \label{fig:coord_schematic}
\end{figure}

The stochastic classical axion field in Eq.~\eqref{eq:classical_axion_field} naturally translates into a stochastic axion gradient, which at the location of the laboratory (which we define as $\mathbf{x}=0$), can be written as \footnote{We have absorbed a phase shift of $\pi/2$ into the definition of $\phi_\mathbf{p}$, which we are at liberty to do since $\phi_\mathbf{p} \to \phi_\mathbf{p} - \pi/2 $ is still uniformly distributed over 2$\pi$.}
\begin{equation}
\boldsymbol{\nabla}a(t) = \sum_\mathbf{p} \sqrt{\frac{2 N_\mathbf{p}}{V \omega_\mathbf{p}}} \cos(p^0 t + \phi_\mathbf{p}) \, \mathbf{p} \,.
\label{eq:axion_gradient_classical}
\end{equation}
Experimentally, the energy shift due to the interaction in Eq.~\eqref{eq:H_int} is measured as the energy shift in spin states of some basis, or equivalently, in spin states along some quantization axis. For the K-$^3$He co-magnetometer introduced in Sec.~\ref{sec:intro} and illustrated in Fig.~\ref{fig:coord_schematic}, we primarily measure the rotation of the atoms' magnetization in the plane of the pump and probe beams, which implies that we are mainly sensitive to energy shifts in states quantized along an axis orthogonal to both pump and probe beams.
Defining this axis as $\hat{\mathbf{m}}$, the gradient of the axion field at discrete times $n\Delta t$ can be understood as a measurable stochastic anomalous magnetic field with a magnitude along $\hat{\mathbf{m}}$ that is given by
\begin{equation}
    \beta_n = \frac{g_\text{aNN}}{\mu} \, \boldsymbol{\nabla}a (n\Delta t) \cdot \hat{\mathbf{m}}(n \Delta t) \,,
    \label{eq:beta_def}
\end{equation}
where $\Delta t$ is the sampling interval,
%
 $n$ is an integer,
and $\mu$ has units of a magnetic dipole moment so that the experimentally measured energy shift from the interaction in Eq.~\eqref{eq:H_int} at time $n\Delta t$ is $\Delta E = \beta_n\, \mu$.

As the authors in~\cite{Lisanti2021} pointed out, because the summation over $\mathbf{p}$ in Eq.~\eqref{eq:axion_gradient_classical} can involve a great many terms and each $\phi_\mathbf{p}$ is, for each distinct value of $\mathbf{p}$, an independent and identically distributed random variable, the set $\{\boldsymbol{\nabla}a (n\Delta t) \,|\, n=0,\ldots,N-1\}$ is a Gaussian process by the Central Limit Theorem. In particular, given a time series $\{0, \Delta t, \ldots, (N-1) \Delta t\}$, the vector $[\boldsymbol{\nabla}a(0),\ldots,\boldsymbol{\nabla}a((N-1) \Delta t)]$, and by extension the vector $[\beta_0, \ldots, \beta_{N-1}]$ that is experimentally measured, follows a multi-variate normal distribution with zero mean (but with a covariance matrix that is not necessarily diagonal) due to the uniformly distributed random phase $\phi_\mathbf{p}$. It is therefore possible to use the statistical properties therein to correctly recover or set upper limits on the axion's coupling strength based on a time series measurement $[\beta_0, \ldots, \beta_{N-1}]$ of the magnetic-like anomalous field $\beta$, as demonstrated in~\cite{Lisanti2021}.

However, it is frequently useful to perform the analysis in the frequency-domain when there are other sources of frequency-dependent noise present. Because the discrete Fourier transform of a time series is an unitary transformation and the normal distribution is closed under linear transformations, one might expect that the Fourier transform of a time series $[\beta_0, \ldots, \beta_{N-1}]$ is still normally distributed. As we show by construction in Appendix~\ref{sec:derivation}, this is indeed true. In the limit where the experimental time span $T$ is much larger than the coherence time $\tau_a$ of the axion, an analytical form of the frequency-domain covariance matrix exists which allows for very efficient computation. Moreover, there is no need here, as required in the time-binned analysis, to re-compute the size of the time bins for different axion masses. For brevity, we leave the full derivation of the frequency-domain covariance matrix to Appendix~\ref{sec:derivation} and only discuss various physical implications of the frequency-domain approach here.

In the non-relativistic limit of Eq. \eqref{eq:axion_gradient_classical}, the oscillation frequency of each axion mode with momentum $\mathbf{p}$ is $p^0 = m_a + m_a v^2/2$, where $v = |\mathbf{p}|/m_a$ is the axion speed. Furthermore, the amplitude of each mode is proportional to the mean occupation number of the mode $N_\mathbf{p}$, which is in turn proportional to the axion's speed distribution $f(v)\,dv$. Consequently, the axion's frequency lineshape is essentially its ``speed spectrum," with each frequency containing contributions from all momentum $\mathbf{p}$ with the same speed. This has been studied in the context of the axion's photon coupling \cite{Foster2018}, but one of the most striking differences of the neutron-spin coupling is that it introduces an additional frequency modulation due to the projection of $\boldsymbol{\nabla}a$ on $\hat{\mathbf{m}}$, the experiment's fixed quantization axis, which rotates at the Earth's angular sidereal frequency $\omega_e$ as the  experiment moves about the Earth's axis.

\begin{figure}[t]
    \centering
    \includegraphics[width=0.49\textwidth]{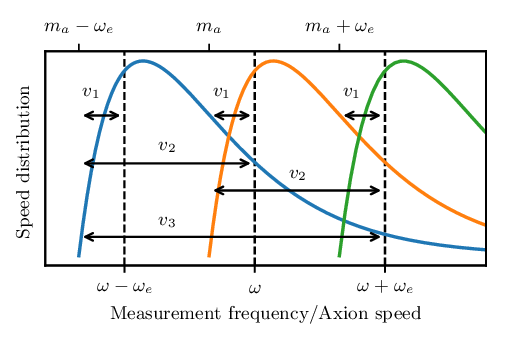}
    \caption{(color online) Correlations between measurement frequencies $\omega$ and $\omega \pm \omega_e$. The original speed distribution (orange), which is zero for all frequencies below the axion mass $m_a$, is down (blue) and up (green) modulated by $\pm \omega_e$. Each frequency measurement (dashed lines) samples axions with varying speeds from both the original and modulated distributions (all denoted with double headed horizontal arrows). For the measurement at $\omega - \omega_e$, only axions with speed $v_1$ can contribute, but axions with speeds $v_1$ and $v_2$ can contribute to the measurement at $\omega$, while the measurement at $\omega + \omega_e$ has contributions from axions with speeds $v_1$, $v_2$, and $v_3$. $\omega - \omega_e$ is correlated with $\omega$ due to the contributions from axions with speed $v_1$, while $\omega$ is correlated with $\omega + \omega_e$ due to contributions from axions with speeds $v_1$ and $v_2$. Similarly, $\omega - \omega_e$ is correlated with $\omega + \omega_e$ because of axions with speed $v_1$.}
    \label{fig:speed_correlation}
\end{figure}

The measured anomalous field $\beta \propto \boldsymbol{\nabla}a\cdot\hat{\mathbf{m}}$ therefore has sidebands spaced $\pm\omega_e$ apart from the frequency of $\boldsymbol{\nabla}a$. If, in addition, $\hat{\mathbf{m}}(t)$ also has a DC-component, i.e. the experiment's quantization axis has a non-zero projection along the Earth's axis, then the frequency spectrum of $\beta$ will also have power proportional to $\boldsymbol{\nabla}a$ at its original frequency. This modulation of the axion's frequency lineshape complicates its speed spectrum since each measured frequency can now also contain contributions with different axion speeds due to sampling from the two additional modulated spectrums. More precisely, the axion speeds that can contribute at any given measured frequency $\omega_k$ are
\begin{equation}
    \aspeed{k}{n} = \sqrt{\frac{2 (\omega_k - (m_a + n \omega_e))}{m_a}}\, ,
    \label{eq:axion_speed}
\end{equation}
for $n\in\{-1,0,1\}$ and $\omega_k - (m_a + n \omega_e) \geq 0$. Accordingly, frequencies that are separated by 1 or 2$\omega_e$ can contain contributions from axions with the same speeds, which means that there can be non-zero correlations between these frequencies since the same phase $\phi_\mathbf{p}$ can be sampled twice in this case. An example of this is shown in Fig. \ref{fig:speed_correlation}.
We emphasize that in the limit of infinite frequency resolution (i.e. the regime where $T \gg \tau_a$) where there is no spectral leakage between neighboring frequency points, only frequencies that are spaced $\pm \omega_e$ and $\pm 2\omega_e$ apart can be correlated. When this limit is not valid, there will be additional correlations between neighboring frequency points, but this correlation is typically small compared to the correlation between frequencies spaced at intervals of $\omega_e$.

\begin{figure}[t]
    \centering
    \includegraphics[width=0.48\textwidth]{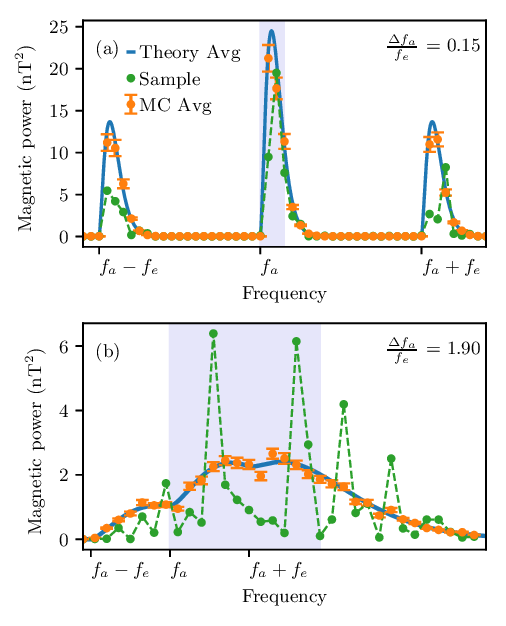}
    \caption{(color online) Magnetic power spectrum for axion linewidths $\Delta f_a$ where (a)~$\Delta f_a/f_e < 1$ and (b)~$\Delta f_a/f_e > 1$. $f_e=\omega_e/2\pi$ is the Earth's sidereal rotation frequency. We have assumed in these plots that the experimental frequency resolution is sufficient to fully resolve the axion's lineshape and that the experimental uncertainty at each point is negligible. The solid blue line is the mean theoretical power spectrum, taken from Eq.~\eqref{eq:R2_avg}, while the orange points give the mean power from 100 Monte-Carlo simulations. Orange error bars give the standard deviation of the mean power from the simulations. The green dots (with connecting lines to guide the eye) show an example of one particular realization of a stochastic axion power spectrum. The data here is generated for a coupling constant of $g_\textrm{aNN}=1$ GeV$^{-1}$. Notice that for the same coupling constant, the peak axion power is lower for (b) compared to (a) due to the broadening of the axion peak for more massive axions. The shaded area denotes the frequency span $[f_a, f_a + \Delta f_a]$.}
    \label{fig:axion_power}
\end{figure}

The width of an axion's spectral line generally depends on its speed distribution. For example, in the Standard Halo Model, the axion's coherence time $\tau_a$ is $\tau_a=1/(\sigma_\textrm{v}^2 m_a)$~\cite{Derevianko2018}, and its linewidth $\Delta f_a$ can be defined to be
\begin{equation}
    \Delta f_a \equiv \frac{1}{\tau_a} = \sigma_\textrm{v}^2 m_a \,.
    \label{eq:linewidth_def}
\end{equation}
Physically, the axion's frequency linewidth is due to its kinetic energy and consequently, more massive axions have larger linewidths because they have more kinetic energy for the same speed. For less-massive axions where $\Delta f_a \ll f_e \equiv \omega_e/2\pi$, the modulation due to the Earth's sidereal rotation splits the expected power spectrum from $\boldsymbol{\nabla}a$ into three peaks. We illustrate this in Fig.~\ref{fig:axion_power}a by plotting the expected magnetic power spectrum (Eq.~\eqref{eq:R2_avg}) from an axion with linewidth $\Delta f_a = 0.15 f_e$ in blue. Orange points show the average power from 100 independent Monte-Carlo simulations of axions with the same mass, while their error bars give the standard deviation of the mean power. Green dots (with dashed connecting lines to guide the eye) give a possible spectrum one might measure for an axion with the same mass. It should be emphasized that in Fig.~\ref{fig:axion_power} we assume a negligible experimental uncertainty, so the scatter of the green points is entirely due to the stochastic nature of the axion signal that one would observe in a particular experimental run.
If the coherence time $\tau_a$ of the axion is sufficiently short so that many independent spectra may be measured during the experiment's data taking run, the averaged power spectrum should (for negligible experimental noise) closely resemble the mean theoretical power spectrum (compare, for instance, the orange points with the blue theoretical line in Fig.~\ref{fig:axion_power}). These conditions apply, for example, to many axion haloscopes \cite{Braine2020, McAllister2017, Zhong2018} that search for axions with masses in the $\mu$eV range that have coherence times on the order of a fraction of a millisecond. However, for extremely light axions with very long coherence times, it may become impractical to obtain many independent spectra (as an example, the coherence time of a 0.1 feV axion is about 75 days long). 
%
In this case, because one particular measured spectrum might differ significantly from the analytical expected magnetic power (see Fig.~\ref{fig:axion_power} for an example), the correct $g_\textrm{aNN}$ cannot be simply extracted by fitting Eq. \eqref{eq:R2_avg} to a measured spectrum. Rather, a more sophisticated analysis should be utilized to correctly extract $g_\textrm{aNN}$ regardless of the stochastic variation of the measured spectrum.

At higher axion masses, $\Delta f_a \propto m_a$ increases and, as shown in Fig.~\ref{fig:axion_power}b, the three peaks re-combine into a single peak when $\Delta f_a > f_e$. Another physical consequence of this broadening is that for the same coupling constant, the peak axion power is lower for a heavier axion than a lighter one since the total axion power must be conserved (for the same dark matter density). This can be observed in Fig.~\ref{fig:axion_power}a and Fig.~\ref{fig:axion_power}b, where both the heavier and lighter axion's spectrum were generated for a fixed coupling constant of $g_\textrm{aNN}$.

\subsection{Likelihood Procedure\label{sec:ll_procedure}}

\begin{figure*}[!ht]
    \centering
    \includegraphics[width=0.98\textwidth]{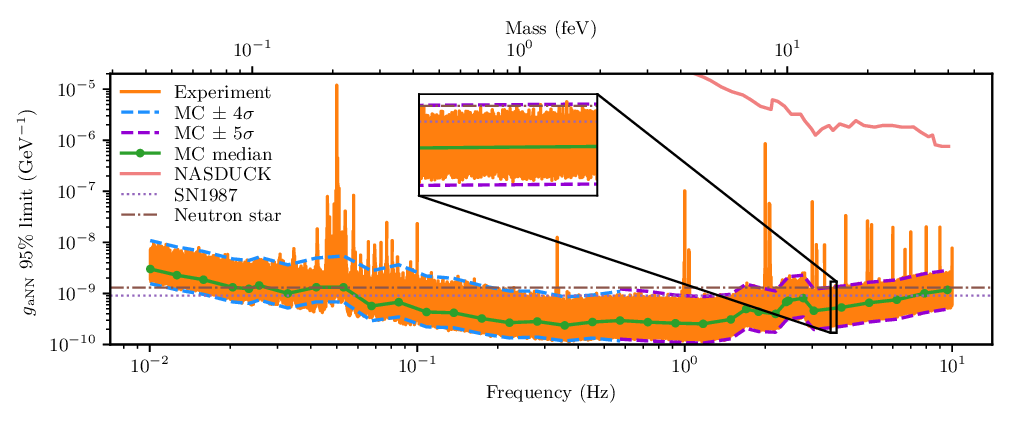}
    \caption{(color online) Orange lines show the 95\% upper limit on $g_\textrm{aNN}$ obtained from our experimental data. The sensitivity of the experiment can be characterized by recovering the 95\% upper limit over an ensemble of Monte Carlo data with no injected signal and plotting the median of those recovered limits (shown as green dots with connecting line to guide the eye). We achieve a median limit of $2.4\times10^{-10}$ GeV$^{-1}$ at 0.36 Hz, approximately five-orders of magnitude stronger than previous laboratory bounds. Due to the large number of axion masses tested, there is a considerable spread in the recovered limits. This is however consistent with the $\pm$4- and 5-$\sigma$ bands (illustrated by pairs of dashed blue and violet lines respectively) obtained via Monte Carlo simulations. At higher axion frequencies, there is a greater density of axion masses on the log-scale plot, but as the magnifying inset shows, the recovered experimental limits are within the $\pm$5$\sigma$ band as expected when viewed at the appropriate scale. The dotted SN1987 limit is a constraint from \cite{Carenza_2019}, which refines the usual one-pion exchange approximation of the nucleon-nucleon bremsstrahlung process. The dash-dot neutron star limit is a 95\% upper limit from \cite{Buschmann2022} that analyzed cooling from five neutron stars. Limits from the NASDUCK collaboration \cite{Bloch2022} are shown as a coral solid line.}
    \label{fig:upper_limits}
\end{figure*}

The likelihood-based analysis used in this work applies to both the regime of low ($\Delta f_a < f_e$) axion mass depicted in Fig.~\ref{fig:axion_power}a, as well as in the high-mass regime of Fig.~\ref{fig:axion_power}b. We now briefly review the likelihood formalism~\cite{Cowan2011} for recovering and/or setting upper limits on the coupling constant $g_\text{aNN}$. The measured experimental spectrum is $\{A_k, B_k \,|\, k=0, \ldots, N-1 \}$, where
\begin{equation}
    A_k = \frac{2}{N}\Re[\tilde{\beta}_k] \,,\, B_k = -\frac{2}{N}\Im[\tilde{\beta}_k]\,,
    \label{eq:Ak_Bk_def}
\end{equation}
and $\tilde{\beta}_k$ is the discrete Fourier transform of $\beta_n$ from Eq.~\eqref{eq:beta_def} given by
\begin{equation}
    \tilde{\beta}_k = \sum_{n=0}^{N-1} \beta_n \exp^{-i\omega_k n \Delta t} \,,\, \omega_k \equiv \frac{2\pi k}{N\Delta t}\,.
    \label{eq:beta_fft_def}
\end{equation}
For a time series with gaps, $A_k$ and $B_k$,  as defined in Eq.~\eqref{eq:Ak_Bk_def}, are determined by performing a linear least-squares fit of the time series to the form
\begin{equation}
    A_k\cos\omega_k t + B_k\sin\omega_k t \, ,
    \label{eq:least_sq_fit_def}
\end{equation}
where $A_k, B_k$ here are the fit parameters. To compute the likelihood of obtaining a spectrum $\{A_k, B_k \,|\, k=0, \ldots, N-1 \}$, we need to have both a signal and background model. The spectrum for an axion with mass $m_a$ can be shown to follow a multi-variate normal distribution with zero mean and a non-diagonal covariance matrix $\boldsymbol{\Sigma}_a(g_{\text{aNN}}, m_a)$ that depends on $g_{\text{aNN}}$ and $m_a$. We assume that the background is white with variance $\sigma_b^2$.  Consequently, the measured signal is normally distributed with zero mean and variance $\boldsymbol{\Sigma} = \boldsymbol{\Sigma}_a + \sigma_b^2 \boldsymbol{\mathbb{1}}$, and the likelihood of measuring the spectrum $\mathbf{d} = \{A_k, B_k \,|\, k=0, \ldots, N-1 \}$ is therefore
\begin{equation}
    L(\mathbf{d}\,|\,g_\text{aNN},\sigma_b) = \frac{1}{\sqrt{(2\pi)^{2N} \det(\boldsymbol{\Sigma})}}
    \exp\left[-\frac{1}{2}\mathbf{d}^T \boldsymbol{\Sigma}^{-1} \mathbf{d}\right].
    \label{eq:likelihood_def}
\end{equation}
Although the experimental noise is frequency dependent, it is to a good approximation white over the bandwidth used for $\mathbf{d}$ (see Table~\ref{tab:likelihood_bandwidth}), except in certain localized cases that result in peaks in the recovered upper limits (see Fig.~\ref{fig:upper_limits} and the discussion in Sec.~\ref{sec:results}).

To set a 95\% upper limit, we define the test statistic
\begin{equation}
    q(g) =
    \begin{dcases*}
    2 \left[\log L(\mathbf{d}|\hat{\mu},\hat{\sigma}_b) - \max_\sigma \log L(\mathbf{d}|g,\sigma)\right]
    &, $\hat{\mu} \leq g$  \\
    0 &, $\hat{\mu} > g$
    \end{dcases*},
    \label{eq:qmu_def}
\end{equation}
where $\hat{\mu}$ and $\hat{\sigma}_b$ are the unconditional maximum likelihood estimators of $g_\text{aNN}$ and $\sigma_b$, respectively:
\begin{align}
    \nonumber
    \hat{\mu} &= \argmax_\mu \log L(\mathbf{d}|\mu, \sigma) \quad\forall\quad \sigma \\
    \hat{\sigma}_b &= \argmax_\sigma \log L(\mathbf{d}|\mu, \sigma) \quad\forall\quad \mu \,.
    \label{eq:uncond_mle_def}
\end{align}
By definition, $q(g) \geq 0$, with larger values of $q(g)$ indicating a greater probability that the hypothesized $g$ is too large and is increasingly incompatible with the data. On the other hand, if the hypothesized $g$ is smaller than the best-fit $\hat{\mu}$, we set $q(g)=0$ because in finding an upper limit, we wish only to set $g$ greater than $\hat{\mu}$, and hence we set $q(g)=0$ when $g < \hat{\mu}$ to indicate that the hypothesized $g$ is not too large. We then find the upper limit by requiring that if the hypothesized $g$ is indeed the true value, then the 95\% upper limit $g_\text{up}$ is such that 95\% of the time, $q(g) \leq q(g_\text{up})$. The cumulative distribution function (c.d.f) of $q(g)$, assuming $g$ is indeed the true value, can be asymptotically shown to be \cite{Wilks1938, Wald1943, Cowan2011}
\begin{equation}
    P(q(g) \leq y) = \Phi(\sqrt{y}) \,,
\end{equation}
where
\begin{equation}
    \Phi(y) = \frac{1}{2}\left[1 + \erf\left(\frac{y}{\sqrt{2}} \right)\right]
\end{equation}
is the c.d.f of a standard normal distribution. Practically, the 95\% upper limit is thus found by numerically solving for $g_\text{up}$ such that $q(g_\text{up}) \approx 2.7055$.

By definition, the test statistic $q(g)$ in Eq.~\eqref{eq:qmu_def} will always set an upper limit above the best-fit value $\hat{\mu}$. To obtain a double-sided confidence interval for a finite best-fit $\hat{\mu}$, we need to use a slightly different test statistic,
\begin{equation}
    t(g) = 2 \left[\log L(\mathbf{d}|\hat{\mu},\hat{\sigma}_b) - \max_\sigma \log L(\mathbf{d}|g,\sigma)\right] \,,
\end{equation}
which is not set to zero when $\hat{\mu} < g$. Rather, by definition, $t(g) \geq 0 $ with larger values of $t(g)$ implying greater incompatibility of the hypothesized value of $g$ with the best-fit $\hat{\mu}$ regardless of whether $g < \hat{\mu}$ or $g > \hat{\mu}$. Assuming that $g$ is indeed the true value, the c.d.f of $t(g)$ is \cite{Cowan2011}
\begin{equation}
    P(t(g) \leq y) = 2\Phi(\sqrt{y}) - 1\,.
\end{equation}
Practically, the higher and lower endpoints of the confidence interval with confidence level CL may therefore be respectively obtained by solving for $g_\text{high} > \hat{\mu}$ such that $P(t(g) < t(g_\text{high})) = (1 - \text{CL})/2$ and $g_\text{low} < \hat{\mu}$ such that $P(t(g) < t(g_\text{low})) = (1 - \text{CL})/2$.

Similarly, to test the significance of the best-fit $\hat{\mu}$, we define the test statistic
\begin{equation}
    q_0 =
    \begin{dcases*}
    2 \left[\log L(\mathbf{d}|\hat{\mu},\hat{\sigma}_b) - \max_\sigma \log L(\mathbf{d}|0,\sigma)\right]
    &, $\hat{\mu} \geq 0$  \\
    0 &, $\hat{\mu} < 0$
    \end{dcases*},
\end{equation}
where we set $q_0 = 0$ when $\hat{\mu} < 0$ to restrict testing for $g_\textrm{aNN} 
\geq 0$ \footnote{Note that $g_\textrm{aNN}$ appears in the likelihood Eq.\eqref{eq:likelihood_def} via the covariance matrix $\boldsymbol{\Sigma}$ as $g_\textrm{aNN}^2$ (see Eq.~\eqref{eq:covAA_def} and \eqref{eq:covAB_def}). Consequently, it suffices to restrict testing to $g_\textrm{aNN} \geq 0$.}. As before, larger values of $q_0$ indicate that the data is increasingly incompatible with the null hypothesis of there being no axion, which may be quantified by the asymptotic c.d.f of $q_0$ (assuming that the true value of $g_\textrm{aNN}$ is indeed 0) \cite{Wilks1938, Wald1943, Cowan2011},
\begin{equation}
    P(q_0 \leq y) = \Phi(\sqrt{y})\,,
    \label{eq:q0_cdf}
\end{equation}
and the asymptotic probability density function (p.d.f) for $q_0$,
\begin{equation}
    f(q_0) = \frac{1}{2}\delta(q_0) + \frac{1}{2\sqrt{2\pi q_0}} e^{-q_0/2} \,,
    \label{eq:q0_pdf}
\end{equation}
where $\delta(q_0)$ is the Dirac delta function. The $p$-value and significance $Z$ of the best-fit $\hat{\mu}$ is thus simply
\begin{equation}
p = 1 - \Phi(\sqrt{q_0}) \quad\textrm{and}\quad Z = \sqrt{q_0} \,.
\label{eq:local_pval_and_Z}
\end{equation}

We now discuss some practical considerations when implementing the above formalism to actual experimental data.
In the validation plots of Appendix~\ref{sec:validation}, we show that our analysis can correctly recover and set upper limits for $g_\textrm{aNN}$ when testing for an axion at the correct mass (i.e. testing for an axion at its actual mass). However, since we do not know the actual mass of the axion, it is necessary to test for it over a large mass parameter space that can in principle span many decades. Given the very different axion lineshapes for a light and heavy axion (see Fig. \ref{fig:axion_power} for example), we expect that testing for an axion at a wrong mass (i.e. testing for an axion at a mass significantly different from its actual mass) should yield a wrong best-fit $g_\textrm{aNN}$ and upper limit. However, it is not immediately obvious how small of a mass/frequency resolution one should use in testing for axions to ensure the recovery of a correct best-fit $g_\textrm{aNN}$ from a real axion with a priori unknown mass.
As we verify in Appendix~\ref{sec:freq_grid}, testing with a spacing of $\approx\Delta f_a/2$ is sufficient to correctly recover or set upper limits on $g_\text{aNN}$ from an axion signal within our mass range of interest.

The need to run the analysis over a large mass parameter space also complicates the interpretation of the significance in recovering a particular best-fit $\hat{\mu}$ with a corresponding discovery test statistic $q_0$. This is because the significance $Z=\sqrt{q_0}$ in Eq.~\eqref{eq:local_pval_and_Z} is only valid when testing at one mass and does not take into account the look-elsewhere effect when testing over a large (e.g., much larger than one) number of axion masses. To relate the global $p$-value, $p_\text{global}$, with a threshold in $q_0$, we observe that by definition, $p_\text{global}$ is the probability that a white noise background will yield $q_0$ that is larger than some threshold $q_\text{th}$ in any of the axion masses tested. For $N$ independent tests, this is equivalent to the complement of the probability that all $N$ tests yield $q_0 \leq q_\text{th}$:
\begin{equation}
    p_\text{global} = 1 - (1 - p)^N \,.
    \label{eq:p_global_def}
\end{equation}
For sufficiently small $p$ so that $(1 - p)^N\approx 1 - N p$, Eq.~\eqref{eq:p_global_def} and \eqref{eq:local_pval_and_Z} readily yield a relation between $q_\text{th}$ and $p_\text{global}$,
\begin{equation}
    q_\text{th} = \left[\Phi^{-1}\left(1 - \frac{p_\text{global}}{N}\right)\right]^2.
    \label{eq:q_thres_def}
\end{equation}

However, as the authors in \cite{Foster2018} noted, the number of independent tests $N$ is frequently smaller than the total number of axion masses tested in the analysis since tests of closely separated axion masses rely on similar experimental spectra and are not independent. Physically, we expect axions that are separated by roughly a linewidth to be independent, with additional modifications due to the experiment's frequency resolution and the sidereal sidebands of the neutron-spin coupling. Consequently, given that the axion's linewidth is due to its kinetic energy, we expect that the axion masses
\begin{equation}
    m_a^{(i)} = m_a^{(0)} (1 + \alpha v_0^2)^i,
    \label{eq:alpha_def}
\end{equation}
for $i=0,\ldots,N - 1$, to be independent, where $v_0$ $\sim$ 220~km/s is the axion's modal speed (see Eq.~\eqref{eq:mb_velocity_dist} and the discussion there) and $\alpha$ is a parameter to be tuned via Monte-Carlo simulations.

\begin{figure}
    \centering
    \includegraphics{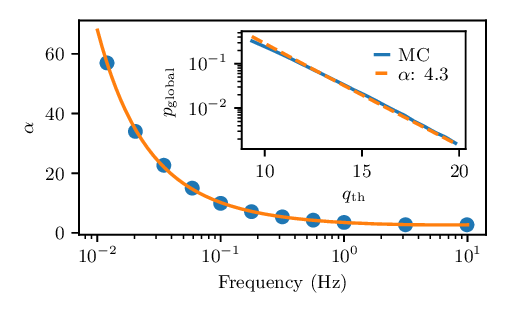}
    \caption{(color online) Main figure: Blue circles give $\alpha$ as computed for that particular frequency. The orange line shows the interpolating function for $\alpha(\nu)$ obtained using the blue circles. Inset: Solid blue line is obtained from 100,000 null Monte Carlo datasets for axion masses centered around 0.562~Hz while the dashed orange line shows the fit of Eq.~\eqref{eq:num_indep_axions_def} to the blue line.}
    \label{fig:look_elsewhere}
\end{figure}

To tune $\alpha$, we generate a large number of null Monte Carlo datasets and perform the likelihood analysis for axion masses between $f_\text{min}$ and $f_\text{max}$ using the same frequency spacing of $\Delta f_a/2$ that is used in the experimental analysis. We then extract $q_\text{th}=\max(q_0)$ from the analysis of each Monte Carlo dataset and obtain the empirical $p_\text{global}$ from the distribution of $q_\text{th}$ over all Monte Carlo datasets. For $N \gg 1$ and $\alpha v_0^2 \ll 1$ (with $v_0$ in $c=1$ units), we obtain from Eq.~\eqref{eq:alpha_def} 
\begin{equation}
    N \approx \frac{1}{\alpha v_0^2} \log\left(\frac{f_\text{max}}{f_\text{min}}\right).
    \label{eq:num_indep_axions_def}
\end{equation}
The tuned value of $\alpha$ can then be obtained by substituting Eq.~\eqref{eq:num_indep_axions_def} into Eq.~\eqref{eq:q_thres_def} and fitting it to the $q_\text{th}$ and $p_\text{global}$ obtained from Monte Carlo simulations. We show an example of such a fit for axion masses centered around 0.562~Hz in the inset of Fig.~\ref{fig:look_elsewhere} where the solid blue line shows $p_\text{global}$ as a function of $q_\text{th}$ as obtained from 100,000 null Monte Carlo datasets, and the dashed orange line is the fit of Eq.~\eqref{eq:q_thres_def} to it.

For lack of computational resources, $[f_\text{min}, f_\text{max}]$ is normally such that $[f_\text{min}, f_\text{max}] \subset \mathcal{F}$, where $\mathcal{F}$ is the set of all axion masses tested in the experimental analysis. The point however, is that once $\alpha$ has been tuned, the number of independent axions $N_\mathcal{F}$ in $\mathcal{F}$ can then be obtained using Eq.~\eqref{eq:num_indep_axions_def} without resorting to Monte Carlo simulations.

As discussed above and in Appendix ~\ref{sec:freq_grid}, we test axions with a spacing of $\Delta f_a/2$ to ensure that we can correctly recover or set upper limits on $g_\text{aNN}$ from an axion signal in our mass range of interest. However, since $\Delta f_a \propto f_a$ from Eq.~\eqref{eq:linewidth_def}, the number of independent axions decreases at lower frequencies since the frequency spacing $\Delta f_a/2$ decreases at low frequencies but the experimental frequency resolution remains constant, which results in a significant overlap of the experimental spectra used in the likelihood analysis at each axion mass. $N_\mathcal{F}$ should therefore be obtained via an integration of Eq.~\eqref{eq:num_indep_axions_def} in logarithmic space
\begin{equation}
    N_\mathcal{F} = \int^{\log\,\max(\mathcal{F})}_{\log\,\min(\mathcal{F})} \,\frac{1}{\alpha(\nu) v_0^2} \, d\nu\,.
    \label{eq:num_indep_axions}
\end{equation}
To obtain $N_\mathcal{F}$, we thus first perform the Monte Carlo procedure above with $f_\text{max}/f_\text{min} \approx 1.0008$ centered around 11 different frequencies between 0.01 to 10 Hz to obtain $\alpha(\nu)$ at 11 discrete points, which is shown as blue circles in Fig.~\ref{fig:look_elsewhere}. We then approximate $\alpha(\nu)$ from 0.01 to 10~Hz using an interpolating function (orange line in Fig.~\ref{fig:look_elsewhere}) and compute $N_\mathcal{F}$ through Eq.~\eqref{eq:num_indep_axions}. For the frequency grid used in our experimental analysis, a $q_0$ of 30.1, 35.6, and 52.1 is estimated by substituting $N_\mathcal{F}$ into Eq.~\eqref{eq:q_thres_def} to correspond to a one-sided global significance of $2\sigma$, $3\sigma$, and $5\sigma$, respectively.

We conclude this section by noting that while the frequency bandwidth used in the calculation of the likelihood in Eq.~\eqref{eq:likelihood_def} at each axion mass must be at least as large as the axion linewidth (including its up and down modulation of $\pm f_e$), its exact value is arbitrary. In general, one desires a bandwidth that is significantly wider than the axion linewidth for proper estimation of the background noise, but is still sufficiently narrow such that the frequency dependent experimental noise is nevertheless approximately white within the chosen bandwidth. Table \ref{tab:likelihood_bandwidth} shows the bandwidth we have used in our analysis.
\begin{table}[!tb]
    \centering
    \renewcommand{\arraystretch}{1.5}
    \begin{tabular*}{0.4\textwidth}{@{\extracolsep{\fill}} cccc}
    \hline\hline
    $f_a$ (Hz) & $\delta f_\textrm{low}$ ($f_e$) & $\delta f_\textrm{high}$ ($f_e$) & $\delta f$ ($f_e$) \\
    \hline
    $(7, 10]$ & 4 & 10 & 14 \\
    $(5, 7]$ & 4 & 8 & 12 \\
    $(2, 5]$ & 4 & 6 & 10 \\
    $[0.01, 2]$ & 4 & 4 & 8 \\
    \hline\hline
    \end{tabular*}
    \caption{Bandwidth $\delta f = \delta f_\textrm{low} + \delta f_\textrm{high}$ spanning frequencies $[f_a - \delta f_\textrm{low}, f_a + \delta f_\textrm{high}]$ that is used to calculate the likelihood for an axion with mass $f_a$. Note that the bandwidth is asymmetrical about $f_a$ since the axion power spectrum is itself asymmetrical due to the fact that the axion's kinetic energy can only be positive. $f_e \approx 11.6$ $\mu$Hz is the frequency of the Earth's rotation with respect to distant stars.}
    \label{tab:likelihood_bandwidth}
\end{table}

\section{Experimental Setup\label{sec:exp_setup}}

The data used in this paper comes from an experiment that was originally designed to search for exotic long-range nuclear spin-dependent forces \cite{Vasilakis2009}, and the interested reader may find more details of the setup in \cite{Vasilakis2011}. In the following, we briefly review the experimental setup of that work and highlight aspects of it that pertain to this study.

At the heart of the experimental setup in \cite{Vasilakis2009} is a K-$^3$He co-magnetometer. Although atomic magnetometers can also be used to measure anomalous magnetic-like fields from an axion's gradient, they are susceptible to ordinary magnetic noise, which limits their usefulness in searching for weak anomalous fields. On the other hand, a K-$^3$He co-magnetometer, which consists of spin polarized K and $^3$He atoms co-located within the same glass cell, can be operated in a way that makes it sensitive to anomalous fields while having a reduced response to ordinary magnetic fields. For the experiment under consideration here, a 2.4 cm diameter spherical aluminosilicate glass cell with 12 amagats of $^3$He, 46 torr of N$_2$ (for quenching excited K atoms during optical pumping), and a droplet of K metal was used. To produce a dense alkali vapor, the cell was placed in a fiberglass oven and electrically heated up to 160$^\circ$C using $\sim$ 200~kHz AC currents that were well above the co-magnetometer's bandwidth. The K was spin polarized by a circularly polarized pump beam from an amplified distributed feedback laser operating at K's $\mathcal{D}_1$ line, and $^3$He was eventually spin polarized through alkali-noble gas spin-exchange collisions with K. A 0.8 mW linearly polarized probe beam blue-detuned 237 GHz from K's $\mathcal{D}_1$ line was directed through the cell perpendicular to the pump beam, and paramagnetic Faraday rotation of its polarization was used to measure the co-magnetometer's signal, which consisted of the projection of K spins along the probe beam. To reduce noise in the probe beam measurement, the probe beam's polarization was first modulated at 50 kHz using a photoelastic modulator together with a quarter-wave plate and was later de-modulated by a lock-in amplifier after detection.

In the absence of a magnetic-like field perpendicular to both the pump and probe beams, the K's magnetization is aligned with the pump beam with zero projection on the probe beam. However, in the presence of such a field, the K's magnetization experiences a torque and rotates into the probe beam, thereby creating an optically detectable signal.
Equivalently, a magnetic-like field along the axis that is perpendicular to both the pump and probe beams results in an energy shift for basis spin states quantized along that sensitive axis (corresponding to $\hat{\mathbf{m}}$ in Eq. \eqref{eq:beta_def}), which results in a rotation of the spin ensemble into the probe beam. The magnitude of this rotation is proportional to the coherence time of the K spins. In the presence of Earth's magnetic field, this coherence time is limited by decoherence from K-K spin exchange collisions. However, at sufficiently low magnetic fields and sufficiently high alkali density, this decoherence can be greatly suppressed~\cite{Happer1973, Happer1977} and significantly improved sensitivity to magnetic-like fields can be achieved. To operate in this low-field regime, and to provide for magnetic shielding from the environment, the cell was placed in five layers of $\mu$-metal shielding with a shielding factor of $\sim10^6$. This allows the K to achieve high sensitivity to both anomalous and ordinary magnetic fields. 

%
%
To suppress the co-magnetometer's sensitivity to ordinary magnetic fields while retaining full sensitivity towards anomalous fields, a bias magnetic field opposite to the sum of the effective magnetic fields from the spin-polarized $^3$He and K atoms is applied \cite{Kornack2002}. Because each spin species in a spherical cell only experiences a magnetic field equal to the sum of the applied bias field and the effective magnetic field of the other spin species (which is anti-parallel to the applied bias field), each spin species only experiences a magnetic field that is equal in magnitude to its own effective magnetic field. Consequently, although the gyromagnetic ratio of $^3$He is much smaller than K due to the much larger mass of the neutron, the resonance frequencies of $^3$He and K are approximately equal in this regime due to the very different magnetic field that each species experiences.
In this regime where the resonance frequencies of both spin species are well matched, both spin ensembles exhibit highly coupled and damped evolution~\cite{Kornack2002}. In particular, the $^3$He magnetization can be shown to adiabatically cancel out small changes in the magnetic field that K experiences so that the co-magnetometer's signal, which is proportional to the projection of the K spins along the probe beam, is to first-order insensitive to ordinary magnetic fields. On the other hand, a slight perturbation of the $^3$He spins due to interactions with a neutron-coupling anomalous field results in a rotation of the $^3$He magnetization that the K experiences, which results in a corresponding perturbation of the K spin along the probe beam that is optically detected.
Meanwhile, an electron-coupling anomalous field couples only to the K spin. This rotates the K's magnetization which is then optically detected so that the K-$^3$He co-magnetometer is, in the final analysis, sensitive to the difference between neutron-coupling ($\beta^n$) and electron-coupling ($\beta^e$) anomalous fields \cite{Kornack2002}.

Assuming that there is no accidental cancellation (i.e., assuming that the axion does not couple to both electrons and neutrons to produce anomalous fields of roughly the same magnitude), we can therefore set limits on neutron and electron coupling magnetic-like anomalous fields independently.
More precisely, given limits on the anomalous magnetic-like field $\beta_\textrm{lim}$, we set $\beta^n < \beta_\textrm{lim}$ and $\beta^e < \beta_\textrm{lim}$, where $\beta^n \propto 0.87\,g_\textrm{aNN}/\mu_\textrm{He}$ and $\beta^e \propto g_\textrm{aee}/\mu_B$ (compare with Eq.~\eqref{eq:beta_def}). $\mu_\textrm{He} \approx 2.148\times10^{-26}$~J/T is here the magnetic dipole moment of $^3$He \cite{Flowers1993} while $\mu_B$ is the Bohr magneton. The factor of 0.87 arises from the fact that the neutron only contributes about 87\% of the nuclear spin of $^3$He~\cite{Friar1990, Ethier2013}. From the above, we note that one can obtain limits on $g_\textrm{aee}$ given limits on $g_\textrm{aNN}$ since $g_\textrm{aee} < 0.87\mu_B\, g_\textrm{aNN}/\mu_\textrm{He}$.

%

For continual suppression of the co-magnetometer's sensitivity to ordinary magnetic fields, the bias magnetic field needs to be periodically adjusted due to slow drifts in the $^3$He magnetization. This is typically performed by an automated routine that minimizes the co-magnetometer's response to a modulated magnetic field along the sensitive axis~\cite{Kornack2002}. Besides zeroing out the co-magnetometer's response to ordinary magnetic fields, this routine was also used to calibrate the co-magnetometer's sensitivity to anomalous fields~\cite{Kornack2002, Vasilakis2011} and was regularly run after every 200~s of continuous data taking. However, since no data can be taken during the zeroing routine, this leads to time gaps between each 200~s record of data that prevents us from obtaining the experimental frequency spectrum by directly performing a Fast Fourier Transform on the time-series data. Rather, we perform a linear least-squares fit of the form in Eq.~\eqref{eq:least_sq_fit_def} to obtain the experiment's frequency spectrum. Due to memory constraints, this was done by first performing the fits over bundles of data that were approximately 24 hours long and storing the best-fit coefficients as well as their fit covariance matrices (assuming a white-noise background) at each fit frequency and for each bundle. The final experimental spectrum was then obtained by computing a coherent weighted average of the best-fit coefficients of each frequency over all the bundles. Before performing our fits, we also filter the data with an appropriate bandpass filter and downsample the original data that was sampled at 200 Hz to a frequency that is at least 4 times larger than the fit frequency. In total, we fitted for this analysis $\sim$ 17 million frequencies from 0.01 to 10~Hz with a frequency resolution of 0.57~$\mu$Hz $\approx$ 1/(40~days).

The original experiment in \cite{Vasilakis2009} consisted of two main data-taking campaigns: one in the spring of 2008 and another in the summer of 2008, with a gap of approximately 50 days in between. Absolute time was recorded in the original experiment as fractional sidereal days since J2000.0 (defined as January 1st, 2000, 12 PM Terrestrial Time), which allows for the orientation of the experiment's sensitive axis to be calculated in terms of Galactic coordinates. Given that the experiment's sensitive axis was oriented vertically throughout both data-taking campaigns, and the coordinates of the experiment at Princeton are $\approx$ 40.35 $^\circ$N, 74.65 $^\circ$W \cite{Justin2011}, the experiment's sensitive axis in Galactic coordinates can be calculated from the definition of a (Greenwich mean) fractional sidereal day and performing a coordinate transformation between the equatorial and Galactic coordinate systems, taking into account precession of the Earth's axis since J2000.0~\cite{McCabe2014}. 

Lastly, we note that throughout the course of the original experiment, the direction of the bias field was periodically flipped as a check on systematic effects. Physically, this causes the K spins to rotate in the opposite direction under the influence of the same anomalous (or magnetic) field pointing along the sensitive axis. Experimentally, this is measured as a sign-flip in the lock-in signal, and we therefore take these field reversals into account by multiplying the co-magnetometer's calibration to anomalous fields with the appropriate signs.

\section{Results\label{sec:results}}

We analyzed our experimental data for roughly 8~million axion frequencies between 0.01 to 10 Hz with a spacing of $\Delta f_a/2$ using the likelihood procedure outlined in Sec.~\ref{sec:freq_domain}. At each axion frequency, a slice of the experimental frequency spectrum with a bandwidth $\delta f$ (tabulated in Table~\ref{tab:likelihood_bandwidth}) was used to calculate the likelihood (Eq.~\eqref{eq:likelihood_def}), which assumes a white noise background. To verify the assumption of a white noise background, we performed a Shapiro-Wilks test on the noise portion of the bandwidth $\delta f$ where the axion's power is expected to be negligible. As can be observed in Fig.~\ref{fig:axion_power}, this can be defined to be frequencies in $\delta f$ that are not within $+2\Delta f_a$ of the axion frequency $f_a$ and its sidebands $f_a \pm f_e$. Based on the $p$-values of the Shapiro-Wilks test, we excluded about 1\% of the $\sim$ 8 million axion frequencies we tested at a significance of 0.0013, corresponding to a one-sided significance of 3$\sigma$. If the noise in our data was completely Gaussian, we would expect a rejection rate of only about 0.1\%; the actual higher rejection rate of 1\% indicates the presence of some non-Gaussian noise in the data.

Since the applicability of our white noise model for frequencies that do not survive the Shapiro-Wilks test is questionable, we do not set limits on $g_\textrm{aNN}$ at these frequencies and they constitute small gaps in our constraints that are not visually discernible in Fig.~\ref{fig:upper_limits} that shows, in orange lines, the remaining 99\% of frequency values at which we set 95\% upper limits on $g_\textrm{aNN}$. None of the frequencies excluded from Fig.~\ref{fig:upper_limits} due to the Shapiro-Wilks test have $q_0$ greater than 52.1 (corresponding to a global significance above 5$\sigma$ if the noise model was actually valid), except for a few frequencies that are nearly an integer multiple of 1~Hz. These peaks feature prominently in Fig.~\ref{fig:upper_limits}, but are unlikely to be true axion signals for reasons that we further elaborate in Sec.~\ref{sec:peaks}.
%
%
To validate our analysis procedure, we also independently calculate in Appendix \ref{sec:exp_validation} the 95\% limits on a subset of experimental data using the time domain method of \cite{Lisanti2021}. Our comparison of both approaches in Fig. \ref{fig:time_freq_exp_compare} indicates that the methods agree well with each other.

Due to the large number of axion frequencies tested, a substantial statistical spread in the computed upper limits is to be expected. This expected statistical spread is illustrated by pairs of dashed blue and violet lines, which respectively give the $\pm$4- and 5-$\sigma$ containment regions of 95\% upper limits obtained by running our likelihood analysis on 10 million null Monte Carlo datasets at multiple frequencies. At each frequency point, the noise used in the Monte Carlo datasets was chosen to accurately reflect the experiment's noise spectrum. The green dots show the median 95\% upper limit at those frequencies, and we achieve a low median limit of $2.4\times 10^{-10}$~GeV$^{-1}$ at~0.36 Hz. Compared to previous laboratory results within the 0.4--40~feV mass range, which come from the original long-range force experiment of this paper \cite{Vasilakis2009} and a more recent CASPEr-ZULF experiment that used low-field NMR techniques to search for astrophysical axions~\cite{Garcon2019}, our results represent an improvement of about five orders of magnitude. More importantly, our Monte-Carlo-validated experimental analysis correctly accounts for the stochastic nature of the interaction in Eq.~\eqref{eq:H_int}, which is crucial for reliably recovering upper limits or best-fit values \cite{Lisanti2021}, but has heretofore not been carefully accounted for in most experimental analyses~\cite{Teng2019, Garcon2019, Jiang2021}. At higher frequencies between 1 and 10 Hz, our constraints are about 2 to 3 orders of magnitude stronger than what was recently reported by the NASDUCK collaboration \cite{Bloch2022}.

Our bounds surpass the upper limit from SN1987A~\cite{Carenza_2019}, which is shown in Fig.~\ref{fig:upper_limits} as a dotted line. This limit from SN1987A is however subject to significant uncertainties due to difficulties in correctly calculating the rate of axion production within the proto-neutron star, as well as complications arising from scattering and absorption within a dense plasma \cite{Raffelt2008, Chang2018, Carenza_2019}. Moreover, if the neutrino emission of SN1987A came not from within the core of a proto-neutron star but from an accretion disk, axion production would not affect the neutrino emission and in that case, constraints on the axion would be invalid~\cite{Bar2020}. Our constraints are also more stringent compared to the neutron star bound from \cite{Buschmann2022} (shown as a brown dash-dot line in Fig.~\ref{fig:upper_limits}), which analyzed the cooling from five neutron stars. Like SN1987A, neutron star cooling arguments also have density-dependent coupling uncertainties \cite{Buschmann2022, Beznogov2018}. In addition, magnetic field decays or other unknown heating mechanisms can plausibly lead to a relaxation of neutron star constraints \cite{Buschmann2022, Beznogov2018}. It is worth noting that there is no universal consensus on the magnitude of the neutron star bound, and it can vary by about an order of magnitude depending on the details of the analysis~\cite{Di_Luzio_2022}. Our results are therefore a useful complementary constraint to both the supernova and neutron star cooling limits.

\begin{figure}[!t]
    \centering
    \includegraphics[width=0.49\textwidth]{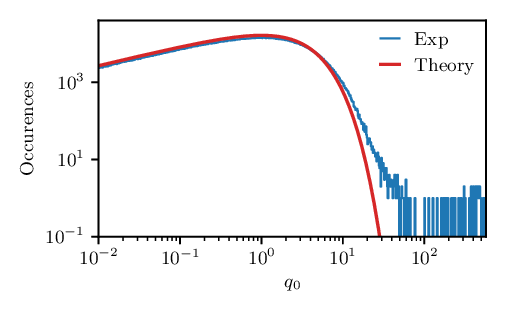}
    \caption{(color online) Distribution of the discovery test statistic is plotted in blue on a logarithmic scale from $10^{-2}$ onward. The red curve shows the expected distribution given the theoretical probability distribution function of $q_0$ for the total number of axion masses tested, assuming that each mass constitutes an independent test. As discussed in the main text, this is only approximately true. A vast majority of the axion masses tested have $q_0 < 0.01$ and are for clarity not shown. Masses with $q_0 > 10$ comprise less than 0.5\% of all masses tested. We do not explicitly account for the look-elsewhere effect in this plot. As we note in the main text, a 5$\sigma$ significance for rejecting the null hypothesis will correspond to a $q_0$ here of 52.1 after taking into account the look-elsewhere effect.
    }
    \label{fig:q0_dist}
\end{figure}

One striking feature of Fig.~\ref{fig:upper_limits} is the presence of multiple peaks in the recovered upper limits that are clearly above the background. This is due to the presence of various peaks in the experiment's power spectrum and can also be seen in the distribution of the $q_0$ discovery test statistic, which we show as a blue line in Fig.~\ref{fig:q0_dist}, where we have restricted the domain to $q_0 \geq 0.01$ for clarity. The expected distribution of $q_0$ (Eq.~\eqref{eq:q0_pdf}) for the total number of axion masses we tested is shown as a red line. At first glance, it might appear that there is an excess of $q_0$ for $q_0 > 10$ and an excessively long tail for $q_0 \gtrsim 50$. However, since not all of our tests are independent due to the finite experimental frequency resolution at low axion masses (see discussion at the end of Sec.~\ref{sec:ll_procedure}), the number of independent excesses is actually smaller than what is suggested by Fig.~\ref{fig:q0_dist}. Moreover, the significance of obtaining a particular value of $q_0$ is no longer given by Eq.~\eqref{eq:local_pval_and_Z} due to the look-elsewhere effect. For example, a significance of $5\sigma$ corresponds to $q_0 = 52.1$ (see Sec.~\ref{sec:ll_procedure}). It is also worth noting that the vast majority of axion masses we tested have $q_0 < 0.01$ and are not shown in Fig.~\ref{fig:q0_dist}. Masses with $q_0 > 10$ comprise less than 0.5\% of the total number of tests performed and there are in total only 63 masses with $q_0 > 52.1$.

Ordinarily, the presence of several masses with $q_0 > 52.1$, corresponding to a significance of more than $5\sigma$ after accounting for the look-elsewhere effect, would warrant further experimental investigation. For example, the bias magnetic field of the co-magnetometer could have been flipped to verify that the effect is not a purely electronic or optical effect, and orienting the experiment differently should in principle also give a different power spectrum from the axion that nevertheless returns the same best-fit $g_\textrm{aNN}$ value from the likelihood analysis. However, since the original experiment was not a dedicated search for astrophysical axions, and since we no longer have active control of the experiment, it is impossible for us to perform experimental checks to rule out spurious signals. Consequently, we are unable to make definitive claims about the origins of these peaks, but we discuss additional analyses below that suggest a non-axion origin for them.

\section{Further analysis of possible axion candidates}
\label{sec:peaks}

\begin{figure}[!t]
    \centering
    \includegraphics{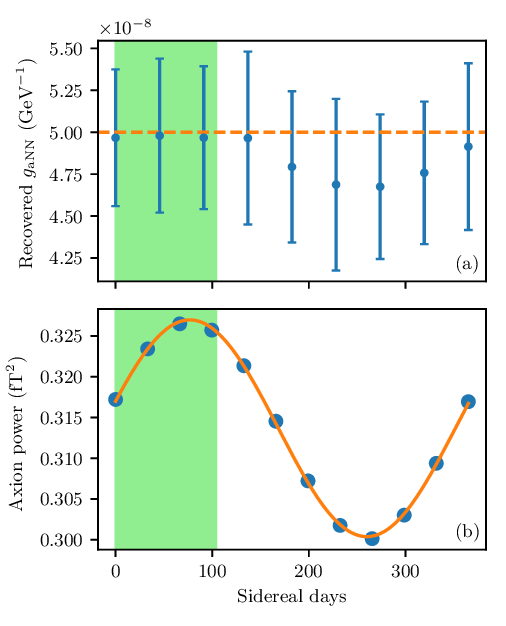}
    \caption{(color online) (a) Expected annual modulation of the \textit{recovered} best-fit $g_\textrm{aNN}$ for a 6 Hz axion. This modulation of the best-fit $g_\textrm{aNN}$ stems from deviation of \textit{both} the magnitude and direction of $\mathbf{v}_E$ from $\langle \mathbf{v}_E \rangle$, where $\mathbf{v}_E$ is Earth's instantaneous velocity relative to the Galactic center and $\langle \mathbf{v}_E \rangle$ is its annual average that we use in our analysis. Because these deviations do not necessarily occur symmetrically, neither is the annual modulation of the best-fit $g_\textrm{aNN}$ always symmetric. This modulation of the \textit{recovered} best-fit $g_\textrm{aNN}$ should be contrasted with the expected annual modulation of the axion's \textit{power} that we show in (b), where blue dots give the expected axion power over 12 months and the orange line is a sinusoidal fit to the blue points. The dashed orange line in (a) indicates the injected $g_\textrm{aNN}$, which is larger than any of the recovered best-fit $g_\textrm{aNN}$ in Fig.~\ref{fig:peaks_ci}. The green shaded window indicates the time span of the experiment, which started data collection on 3022.63 sidereal days since J2000.0. As the results in (a) show, no significant deviation of the recovered $g_\textrm{aNN}$ is expected over the time span of our experiment due to our use of $\langle\mathbf{v}_e\rangle$ rather than the instantaneous $\mathbf{v}_E$ in our analysis. Vertical lines in (a) give the 1-$\sigma$ containment interval for the best-fit $g_\textrm{aNN}$ from 100 Monte Carlo datasets. The simulated data has a white noise background of 1.6 fT/$\sqrt{\textrm{Hz}}$, which is comparable to the experiment's noise level.}
    \label{fig:annual_modulation}
\end{figure}

In this section, we further analyze possible axion candidates, which we here define as those axion masses with $q_0 > 52.1$. As discussed in Sec.~\ref{sec:exp_setup}, data for the original experiment was taken over spring and summer of 2008, with a gap of approximately 50 days in between both campaigns during which systematic improvements particular to the original experiment were performed. A true axion signal should nevertheless persist in both datasets, and we would expect that the recovered best-fit $g_\textrm{aNN}$ from both analyses would not be significantly different from each other. This is true even though we only use an average Earth velocity $\langle\mathbf{v}_E\rangle$ in our analysis of both datasets instead of Earth's instantaneous velocity $\mathbf{v}_E$. As we show in Fig.~\ref{fig:annual_modulation}a, the expected discrepancy between the recovered best-fit $g_\textrm{aNN}$ due to this approximation is negligible over the duration of our experiment. Consequently, any candidate that has significantly different best-fit $g_\textrm{aNN}$ between the spring and summer datasets is likely to not be a true axion.

\begin{figure*}[!ht]
    \centering
    \includegraphics[width=0.98\textwidth]{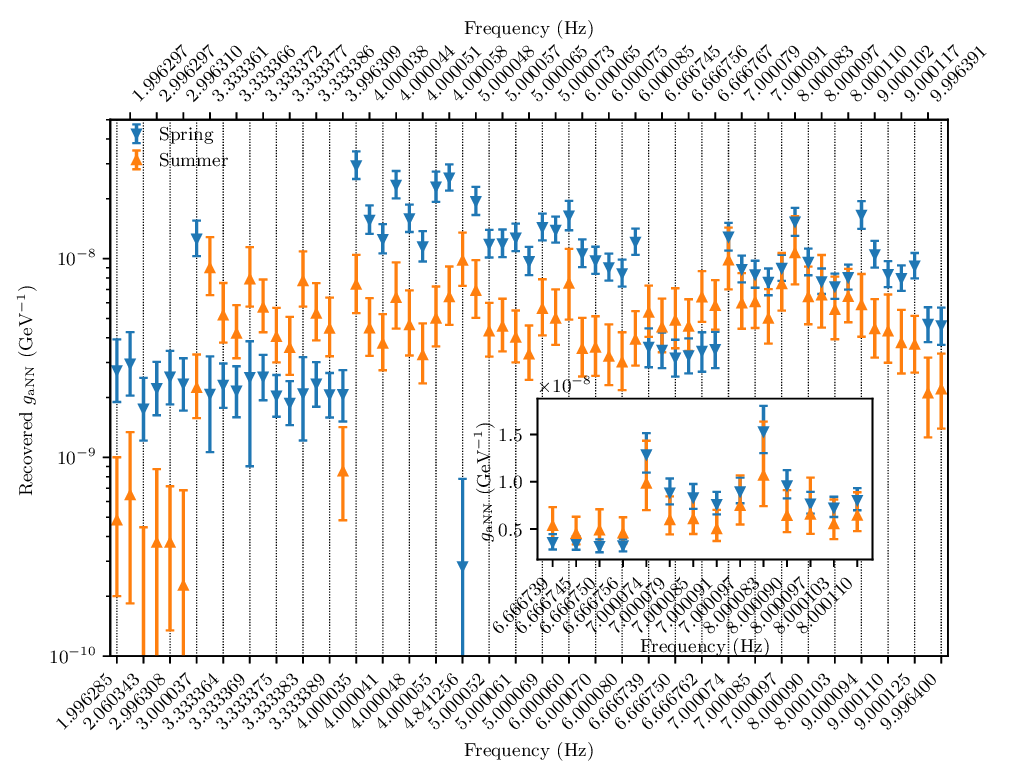}
    \caption{Best-fit $g_\textrm{aNN}$ recovered from both the spring and summer datasets for all axion masses with $q_0 > 52.1$ corresponding to a significance of more than $5\sigma$ after taking into account the look-elsewhere effect. Error bars show the 95\% confidence intervals for the best-fit values. Of the 63 possible candidates shown here, only 14 have overlapping confidence intervals. These candidates are also shown in the inset.}
    \label{fig:peaks_ci}
\end{figure*}

In Fig.~\ref{fig:annual_modulation}a, the central value of each data point gives the average best-fit $g_\textrm{aNN}$ over 100 Monte Carlo datasets with a simulated data-taking interval of a week for a 6~Hz axion. Vertical lines indicate the 1$\sigma$ containment interval of those best-fit values, which is a function of both the signal-to-noise ratio and the frequency resolution of the experiment relative to the axion linewidth. The latter is due to the stochastic nature of the astrophysical axion signal where it is necessary to sample the axion signal over several coherence times to have a more reliable measure of $g_\textrm{aNN}$. Due to the directional nature of the neutron-spin coupling, our analysis is sensitive to \textit{both} the magnitude and direction of $\mathbf{v}_E$. Since both of these quantities do not necessarily deviate symmetrically from their average values, this modulation of the best-fit $g_\textrm{aNN}$ stemming from our use of $\langle\mathbf{v}_E\rangle$ rather than $\mathbf{v}_E$ is therefore also not always symmetric. This asymmetric modulation of the \textit{recovered} best-fit $g_\textrm{aNN}$ should be contrasted with the symmetric modulation of the expected raw axion power which we show in Fig.~\ref{fig:annual_modulation}b.

As an aside, it is interesting to note here that the annual peak-to-peak \textit{power} modulation of thermalized axions in the Standard Halo Model is only about 8\% of its mean value, making this a relatively small effect that will be especially difficult to observe for low-frequency axions with coherence times on the order of a year.

In Fig.~\ref{fig:annual_modulation}a, we chose to simulate a 6~Hz axion because it is close in frequency to several of the axion candidates that we discuss below. Moreover, we inject the axion signal with a relatively large $g_\textrm{aNN}$ of $5\times10^{-8}$ GeV$^{-1}$ (shown as a dashed orange line in Fig.~\ref{fig:annual_modulation}a) and for a sufficiently long measurement time so that the 1$\sigma$ containment interval would be comparable to the experimental confidence intervals for our best-fit $g_\textrm{aNN}$. Nevertheless, the expected modulation of the recovered $g_\textrm{aNN}$ in Fig.~\ref{fig:annual_modulation}a is still negligible over the duration of our experiment, which is demarcated in the figure by the green shaded window. Consequently, we would expect the 95\% confidence intervals of the recovered best-fit $g_\textrm{aNN}$ values from both the spring and summer datasets to overlap with each other if they are indeed due to a true axion signal.

In Fig.~\ref{fig:peaks_ci}, we plot the best-fit $g_\textrm{aNN}$ recovered from both the spring and summer datasets for all axion masses with $q_0 > 52.1$ in the full combined dataset. Vertical lines indicate their 95\% confidence intervals obtained by analyzing the data in each season using the formalism in Sec.~\ref{sec:ll_procedure}. Although there are numerous prominent peaks in Fig.~\ref{fig:upper_limits} with frequencies below 2 Hz, their significance does not exceed 5$\sigma$ because the width of their peaks is generally much broader than the expected axion linewidth at those frequencies. It therefore turns out that there are only 63 axion candidates above $\approx$ 2 Hz that have significance above $5\sigma$ after taking into account the look-elsewhere effect. Out of these 63 possible axion candidates with $q_0 > 52.1$, only 14 have overlapping 95\% confidence intervals. We deem the other 49 axion candidates with non-overlapping $g_\textrm{aNN}$ confidence intervals as unlikely to be true axions due to inconsistencies from both the spring and summer datasets, and we do not perform any further analysis on them. For the remaining 14 axion candidates with overlapping 95\% confidence intervals, we further perform a peak shape analysis on them that we describe below.

\begin{figure}[!t]
    \centering
   \includegraphics[width=0.49\textwidth]{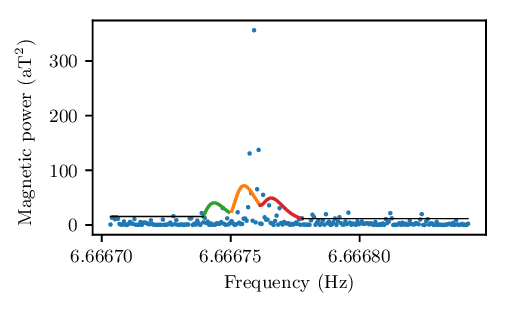}
   \caption{(color online) Experimental magnetic power near 6.666~Hz (dots). The average magnetic power of a signal candidate identified by the likelihood analysis at 6.66675~Hz is shown by the solid line. For the purposes of testing an axion candidate directly against the signal model (rather than against a white-noise null hypothesis as in our likelihood analysis), we split the signal power into three regions identified by colored lines to calculate the power under each peak. These peaks are due to sideband modulations of the axion's power due to Earth's rotation about its axis (see discussion around Eq.~\eqref{eq:axion_speed}). The selection of these three regions is driven by a need to discriminate between true and false axion signals (see discussion in Sec.~\ref{sec:peaks} of the text and Fig.~\ref{fig:peakprob}).}
   \label{fig:axionpeak666}
\end{figure}

A unique feature of the signal lineshape in the axion frequency range that we examined is the appearance of 3 distinct peaks spaced by the Earth's sidereal rotation frequency. This allows us to discriminate between sidereal modulation of the axion signal and monochromatic signals that are likely due to terrestrial sources. Figure~\ref{fig:axionpeak666} shows an example of an experimental peak at 6.666~Hz where the likelihood analysis identifies several possible axion candidates with $q_0$ of around 200. However, because the likelihood analysis is effectively a hypothesis test for the axion model against the null hypothesis of a white noise background, any relatively narrow deviation from a flat spectrum results in a high $q_0$ value favoring the axion hypothesis even if the lineshape match to a true axion signal is poor.

\begin{figure}[!t]
    \centering
   \includegraphics[width=0.49\textwidth]{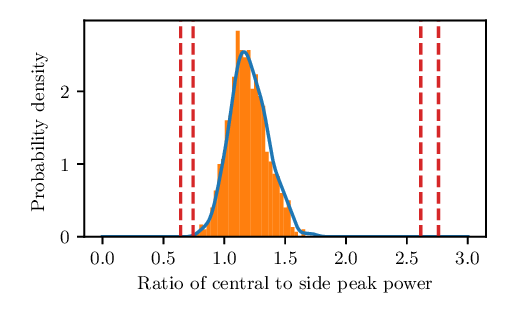}
   \caption{(color online) A histogram of ratios of the power in the central axion peak (colored orange in Fig.~\ref{fig:axionpeak666}) relative to the average power of the side peaks (colored green and red in Fig.~\ref{fig:axionpeak666}) based on Monte-Carlo simulations for a signal candidate at 6.66675~Hz. From the Monte-Carlo simulations, this ratio should be centered around approximately 1.2 for a true axion signal. The solid blue line shows a smoothed distribution of the ratio. Red lines indicate the experimental ratios for the four peaks near 6.666 Hz shown on the inset of Fig.~\ref{fig:peaks_ci}. Given their distance from the Monte-Carlo distribution, we can conclude with $>$ 90\% confidence that these peaks are not true axions.}
     \label{fig:peakprob}
\end{figure}

An obvious way to directly test an axion candidate against the signal model is to fit its lineshape to the expected lineshape of an axion. Nevertheless, fitting the lineshape of the expected axion signal is in general difficult since it has large statistical fluctuations in each frequency bin. To reduce this uncertainty, one has to combine the power in many bins together, but the exact number of bins to combine depends on the axion linewidth which is itself proportional to the axion frequency. In order to have a general lineshape analysis, we calculate the total power under each of the three expected axion peaks. When the peaks overlap, we separate the signal into three frequency bands as shown for example by the different colors in Fig.~\ref{fig:axionpeak666}. We then calculate the  ratio of the central peak power to the average of the two side peak powers. From Monte-Carlo simulations of axions in our mass range of interest, this ratio is generally expected to be slightly larger than 1 for a good axion candidate. Physically, this ratio depends on the projection of the experiment's sensitive axis (defined as $\hat{\mathbf{m}}$ in Eq.~\eqref{eq:beta_def}) on the axis of Earth's rotation. An analytical result of this ratio can be obtained for the expected magnetic power spectrum (see Eq.~\eqref{eq:R2_avg} and Table~\ref{tab:sensitive_axis_comps}), but there is some variability for a single measurement due to the stochasticity of dark matter axions, which requires additional verification via Monte-Carlo simulations. In Fig.~\ref{fig:peakprob}, we plot a distribution of relative peak power ratios based on 1000 Monte-Carlo simulations for a candidate signal at 6.66675~Hz. We also show the experimental power ratios for the four signal candidates near 6.666~Hz that have overlapping $g_\textrm{aNN}$ confidence intervals from both summer and spring datasets. One can see that all signal candidates are excluded with $>90\%$ probability. Similar analyses are performed for other candidate peaks near 7~Hz and 8~Hz, and they are similarly inconsistent with the expected axion lineshape.

Lastly, we note that the 6.666 Hz candidates are an integer multiple of the 3.333 Hz candidates which have non-overlapping confidence intervals in Fig.~\ref{fig:peaks_ci}, and that the 7 and 8 Hz candidates are integer multiples of a 1~Hz peak that is likely due to a clock signal in one of our electronics. Given the suspicious coincidences of these candidates and the fact that their lineshapes do not agree well with a true axion's lineshape, we do not in the final analysis deem any of them to be serious axion contenders.

\section{Conclusion\label{sec:conclusion}}

Axions are well-motivated dark matter candidates that arise in many theories beyond the Standard Model. In this paper, we re-analyzed approximately 40 days of data from a K-$^3$He co-magnetometer that was originally built for a long-range force experiment to search instead for dark matter axions coupling to the neutron spin of $^3$He.

Compared to long-range force experiments, searches for dark matter axions produce energy shifts that are only suppressed by one factor of the small coupling constant $g_\textrm{aNN}$ but their study is complicated by interference effects that result in stochastic experimental signatures. To correctly account for this stochastic signal, we developed a likelihood analysis to analyze the signal of the K-$^3$He co-magnetometer in the frequency domain from 0.01 Hz to 10 Hz ($\approx$ 0.04 feV to 40 feV). Assuming that axions comprise all of the dark matter in the Solar neighborhood, we were able to constrain $g_\textrm{aNN} < 2.4\times 10^{-10}$~GeV$^{-1}$  (median 95\% confidence level) for axion masses between 0.4 to 4 feV. At higher masses (frequencies), the co-magnetometer's sensitivity to anomalous fields is limited by the resonance frequency of $^3$He (about 20 Hz in the original experiment). It is in principle possible to extend the analysis for lower frequencies, but the co-magnetometer's sensitivity also deteriorates at lower frequencies. This is typically due to low frequency noise from slow drifts in the pump and probe beams.

Our limits represent a significant five orders-of-magnitude improvement over previous laboratory bounds and serve as a useful verification of astrophysical constraints that have comparable limits but are subject to substantial uncertainties.
Moreover, as discussed in Sec.~\ref{sec:exp_setup}, the K-$^3$He co-magnetometer is also, barring an unlikely accidental cancellation, sensitive to anomalous fields coupling to the electron spin so that analogous constraints on the axion-electron coupling $g_\textrm{aee}$ can be easily obtained from Fig.~\ref{fig:upper_limits} and multiplying the limits by a simple re-scaling factor of 0.87$\mu_B/\mu_\textrm{He}$ (see Sec.~\ref{sec:exp_setup} for more details).
%

Peaks in the magnetic power spectrum of the original experiment resulted in several persistent possible axion candidates with significance greater than 5$\sigma$ after taking into account the look-elsewhere effect. Their high significance suggests that the null hypothesis of a white noise background should be rejected, but additional verification is required before acceptance of an axion hypothesis. Since we no longer have active control of the experiment, we are unable to perform detailed experimental checks on these candidates. However, analysis of their lineshapes shows significant deviation from their expected values, which suggests a non-axion origin for all of them.

\begin{acknowledgements}
We thank M. Moschella for helpful discussions and for his assistance in validating some aspects of the analysis. JL, WT, and MR were supported by Simons Foundation award number 641332. ML is supported by the DOE under Award Number DE-SC0007968.  This work was performed in part at the Aspen Center for Physics, which is supported by NSF grant PHY-1607611.   The work presented in this paper was performed on computational resources managed and supported by Princeton Research Computing, a consortium of groups including the Princeton Institute for Computational Science and Engineering (PICSciE) and the Office of Information Technology's High Performance Computing Center and Visualization Laboratory at Princeton University.
\end{acknowledgements}

\appendix

\section{Validation of frequency domain likelihood analysis\label{sec:validation}}

We introduced our frequency domain likelihood analysis in Sec.~\ref{sec:freq_domain} and provide a full derivation of the frequency domain covariance matrix in Appendix~\ref{sec:derivation}. Here, we present our validation of the frequency domain analysis by comparing it with the time domain analysis recently published in \cite{Lisanti2021}. We first compare both approaches on the same set of Monte Carlo data, before testing them on the same subset of experimental data.

\subsection{Validation on Monte Carlo data}

\begin{figure}[!t]
    \centering
    \includegraphics[width=0.48\textwidth]{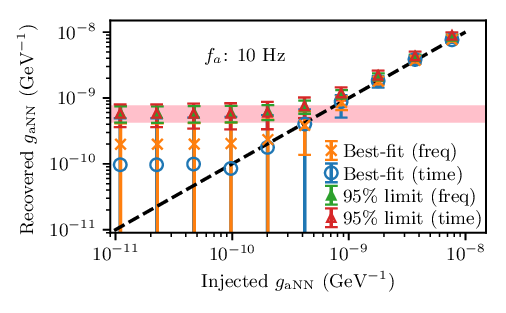}
    \caption{
    (color online) Signal injection and recovery for an axion mass of 10 Hz. To create these plots, we generated 100 Monte Carlo time-series datasets sampled at 50 Hz and each lasting 40 days long with a white noise background of 900 aT/$\sqrt{\textrm{Hz}}$. The time-binned analysis in \cite{Lisanti2021} and the frequency domain analysis in Sec.~\ref{sec:freq_domain} were then both used to recover the best-fit and 95\% upper limit of the coupling constant. Error bars are the standard deviation of the recovered best-fit and upper limits values over the 100 simulated data sets while the markers denote the mean best-fit and upper limits over the data sets. The null limit, found by running the analysis on 100 null Monte Carlo data sets, is shown as a pink band.
    }
    \label{fig:mc_injection_recovery}
\end{figure}

In Fig.~\ref{fig:mc_injection_recovery}, we show the signal recovery plot of 100 Monte Carlo time-series datasets that were each 40 days long. The simulated data sets were injected with a constant white noise background and a simulated axion signal of varying strengths spanning a few decades. The time-binned analysis in \cite{Lisanti2021} and the frequency domain analysis were then independently applied to the same simulated data to recover the best-fit $g_\textrm{aNN}$ (i.e. the unconditional maximum likelihood estimator of $g_\textrm{aNN}$) and 95\% upper limits shown in Fig.~\ref{fig:mc_injection_recovery}. We emphasize that the Monte Carlo data here was generated in the time domain and is completely independent of the frequency domain likelihood formalism. It is thus a non-trivial validation of the methodology presented in Appendix~\ref{sec:derivation}.

\begin{figure}[!t]
    \centering
    \includegraphics[width=0.49\textwidth]{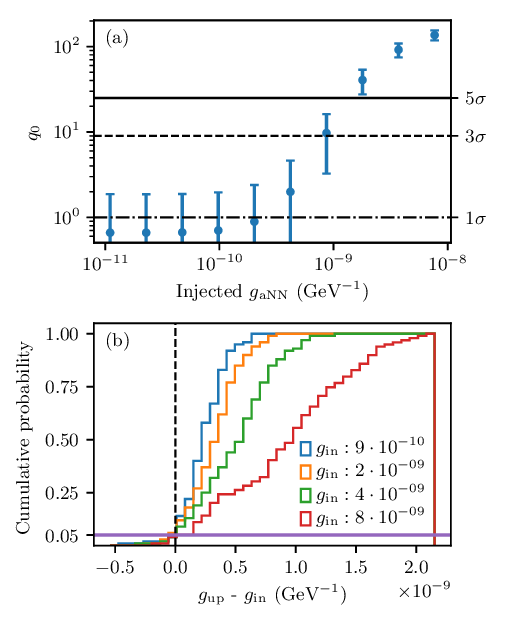}
    \caption{(color online) (a) $q_0$ as a function of injected $g_\textrm{aNN}$ for the case of a 10 Hz axion. The marker denotes the average of the $q_0$ test statistic recovered from 100 independent Monte Carlo simulations, while the error bars denote the corresponding standard deviation. Horizontal lines with labels of 1, 3, and 5$\sigma$ on the right indicate the significance of obtaining a particular value of $q_0$ (see Sec.~\ref{sec:ll_procedure} for more details). (b) Empirical cumulative distribution of the recovered 95\% upper limit $g_\textrm{up}$ for the case of a 10 Hz axion where the injected $g_\textrm{in}$ is above the null limit (see Fig.~\ref{fig:mc_injection_recovery}). As expected, the recovered limits are below the injected $g_\textrm{in}$ for $\approx$ 5\% of the time.}
    \label{fig:mc_q0_ulim}
\end{figure}

The unconditional maximum likelihood estimator of $g_\textrm{aNN}$ in Fig.~\ref{fig:mc_injection_recovery} correctly recovers the injected value for a sufficiently strong injected signal. However, when the injected signal falls below the noise floor, the method is, as expected, unable to reliably recover the injected $g_\textrm{aNN}$ as seen in the diverging error bars, which shows the standard deviation of the best-fit $g_\textrm{aNN}$ recovered over all 100 data sets. This can also be seen in Fig.~\ref{fig:mc_q0_ulim}a, which shows the test statistic $q_0$ obtained via the frequency-domain method as a function of the injected $g_\textrm{aNN}$ for the 10 Hz axion. At low values of injected $g_\textrm{aNN}$, the significance of the recovered best-fit value is $\sim 1\sigma$, indicating that the null hypothesis of no axion should be preferred. However, at larger injected $g_\textrm{aNN}$ values, the test statistic and significance of the recovered best-fit value increases accordingly before eventually saturating in the limit of high signal-to-noise ratio.

Although we cannot reliably recover the best-fit value when the injected signal disappears below the noise floor, we can nevertheless still reliably set upper limits. This can be seen for both the time and frequency domain analysis in Fig.~\ref{fig:mc_injection_recovery} where both methods set the 95\% upper limits in the low injected $g_\textrm{aNN}$ regime with error bars that agree well with the null limit that is shown as a pink band. At higher injected $g_\textrm{aNN}$ above the null limit, the logarithmic scale of Fig.~\ref{fig:mc_injection_recovery} makes it difficult to discern if the upper limits are recovered correctly. To validate the upper limits in this regime, we plot, for the 10 Hz axion, the cumulative distribution functions of the recovered 95\% upper limits $g_\textrm{up}$ for all injected $g_\textrm{in}$ above the null limit and compare it against the actual injected $g_\textrm{in}$. As Fig.~\ref{fig:mc_q0_ulim}b shows, the recovered upper limits are indeed only below $g_\textrm{in}$ about 5\% of the time, which indicates that the analysis is working as intended.

\subsection{Validation on experimental data\label{sec:exp_validation}}

\begin{figure}[!t]
    \centering
    \includegraphics[width=0.49\textwidth]{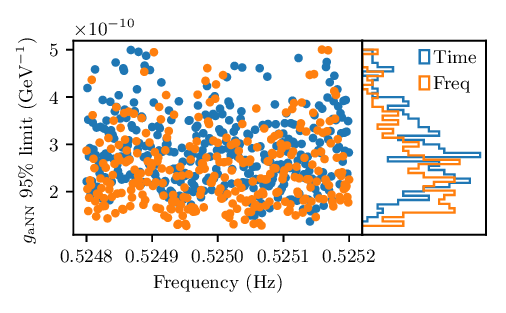}
    \caption{(color online) Left: Comparison of the 95\% upper limit obtained via the frequency domain approach of this paper (orange) and the time domain approach of \cite{Lisanti2021} (blue) over a small frequency range of the experimental data in the vicinity of 0.525 Hz. Note that the calculations were completely independent and that there is some misalignment of the frequency grids. Right: Histogram of the obtained limits in the frequency range of the left panel.}
    \label{fig:time_freq_exp_compare}
\end{figure}

Besides checking that both approaches work on the same Monte Carlo datasets, we have also checked that both methods give similar results on a subset of the actual experimental data. In Fig. \ref{fig:time_freq_exp_compare}, we use both the frequency domain approach of this paper and the time domain approach of \cite{Lisanti2021} to independently calculate the 95\% upper limit on $g_\textrm{aNN}$ from the experimental data over a small frequency range in the neighborhood of 0.525~Hz where the experimental noise spectrum is relatively white. As the scatter plot and histogram of Fig. \ref{fig:time_freq_exp_compare} show, both methods agree with each other quite well.

\section{Stochastic properties of the axion gradient's frequency spectrum\label{sec:derivation}}

Throughout this paper, we have claimed that the experimental frequency spectrum $\mathbf{d} = \{A_k, B_k \,|\, k=0,\ldots,N-1\}$ (see Sec.~\ref{sec:freq_domain}), where $A_k$ and $B_k$ are as defined in Eq. \eqref{eq:Ak_Bk_def}, is normally distributed with zero mean and has a non-diagonal covariance matrix $\Sigma_a$. These stochastic properties were then used to calculate the likelihood of measuring a particular frequency spectrum. We now prove that $\mathbf{d}$ is indeed normally distributed with zero mean and provide the full derivation of $\Sigma_a$ in this appendix.

Before obtaining the stochastic properties of $A_k$ and $B_k$ however, we note that the stochastic axion gradient, as written in Eq. \eqref{eq:axion_gradient_classical}, is a sum over all 3-momentum $\mathbf{p}$ of the axion, which have different directions and magnitudes. However, since the axion's frequency spectrum can only depend on the magnitude of its momentum, it is clear that we need to first integrate out the angular degrees of freedom of $\mathbf{p}$ in Eq. \eqref{eq:axion_gradient_classical} before we can obtain the stochastic properties of $A_k$ and $B_k$.

\subsection{Integrating out the angular dependence in $\mathbf{p}$\label{sec:deriv_integration}}

We do this by first re-writing the sum over all $\mathbf{p}$ in Eq. \eqref{eq:axion_gradient_classical} to a sum over infinitesimal domains $\Omega_\mathbf{p}$. Within each infinitesimal $\Omega_\mathbf{p}$ domain, we may safely assume that $\mathbf{p}$, $N_\mathbf{p}$, and $\omega_\mathbf{p}$ are approximately the same. However, because $\phi_\mathbf{p} \sim U(0, 2\pi)$ is a random number, it cannot be assumed to be constant within any infinitesimal domain and therefore requires an additional sum
\begin{align}
\nonumber
\boldsymbol{\nabla}a(t) &= \sum_{\Omega_\mathbf{p}} \sqrt{\frac{2 N_\mathbf{p}}{V \omega_\mathbf{p}}} \sum_{\mathbf{k}\in\Omega_\mathbf{p}} \cos(p^0 t + \phi_\mathbf{k}) \mathbf{p} \\
\nonumber
&= \sum_{\Omega_\mathbf{p}} \sqrt{\frac{2 N_\mathbf{p}}{V \omega_\mathbf{p}}}
\left(\cos p^0 t \sum_{\mathbf{k}\in\Omega_\mathbf{p}} \cos \phi_\mathbf{k} \right. \\
\nonumber
& \hspace{3cm} \left.- \sin p^0 t \sum_{\mathbf{k}\in\Omega_\mathbf{p}} \sin \phi_\mathbf{k} \right) \mathbf{p} \\
\nonumber
&= \sum_{\Omega_\mathbf{p}} \sqrt{\frac{2 N_\mathbf{p}}{V \omega_\mathbf{p}}}
\left(x_\mathbf{p}\cos p^0 t - y_\mathbf{p} \sin p^0 t \right) \mathbf{p} \\
&= \sum_{\Omega_\mathbf{p}} \sqrt{\frac{N_\mathbf{p} M_\mathbf{p}}{V \omega_\mathbf{p}}} \alpha_\mathbf{p}
\cos(p^0 t + \Phi_\mathbf{p}) \mathbf{p}.
\label{eq:axion_gradient_classical2}
\end{align}
In going from the second to third line, we have noted that each $\cos\phi_\mathbf{k}$ and $\sin\phi_\mathbf{k}$ are identically and independently distributed. Consequently, by the Central Limit Theorem, in the limit of $M_\mathbf{p}\to\infty$, where $M_\mathbf{p}$ is the number of modes in $\Omega_\mathbf{p}$, $\sum_{\mathbf{k}\in\Omega_\mathbf{p}} \cos\phi_\mathbf{k} \to x_\mathbf{p} \sim N(M_\mathbf{p}\Evalue[\cos\phi_\mathbf{k}], M_\mathbf{p}\Var[\cos\phi_\mathbf{k}]) = N(0, M_\mathbf{p}/2)$. Similarly, $\sum_{\mathbf{k}\in\Omega_\mathbf{p}} \sin\phi_\mathbf{k}$ converges to $y_\mathbf{p}\sim N(0, M_\mathbf{p}/2)$. Moreover, since $\Evalue[\cos\phi_\mathbf{k}\sin\phi_\mathbf{k}] = \int_0^{2\pi} d\phi \cos\phi\sin\phi/2\pi = 0$, the two sums are independent and $x_\mathbf{p}$ and $y_\mathbf{p}$ are therefore independent, normally distributed variables with variance $M_\mathbf{p}/2$. Thus, the last line follows, where we have re-scaled and performed a change of variables to Rayleigh distributed $\alpha_\mathbf{p}\sim R(1)$ and uniformly distributed $\Phi_\mathbf{p}\sim U(0, 2\pi)$. Furthermore, we observe that since $N_\mathbf{p}$ is the mean occupation number of each mode in $\Omega_\mathbf{p}$, and $M_\mathbf{p}$ is the number of modes in $\Omega_\mathbf{p}$, their product is the mean number of axions in $\Omega_\mathbf{p}$. If we assume that axions make up the entirety of the local dark matter density $\rho_\textrm{DM}\approx 0.3$ GeV/cm$^3$ \cite{Catena2010}, and that they have a momentum distribution $f(\mathbf{p})\, \textrm{d}\mathbf{p}$ such that $\int \textrm{d}\mathbf{p}\, f(\mathbf{p}) = 1$, then we may write $N_\mathbf{p} M_\mathbf{p} = \rho_\textrm{DM} V f(\mathbf{p}) (\Delta p)^3/\omega_\mathbf{p}$, which gives
\begin{align}
\nonumber
\boldsymbol{\nabla}a(t) &= \sum_{\Omega_\mathbf{p}} \frac{\sqrt{\rho_\textrm{DM} f(\mathbf{p}) (\Delta p)^3}}{\omega_\mathbf{p}}
\alpha_\mathbf{p} \cos(p^0 t + \Phi_\mathbf{p}) \mathbf{p} \\
\nonumber
&\approx \sum_{\Omega_\mathbf{p}} \sqrt{\rho_\textrm{DM} f(\mathbf{v}) (\Delta v)^3} \, \alpha_\mathbf{v} \\
& \hspace{2cm} \times \cos\left(m_a \left(1 + \frac{1}{2}v^2\right)t + \Phi_\mathbf{v}\right)\mathbf{v}, 
\end{align}
where in the last line we have taken the non-relativistic limit since halo axions are bound within the galaxy and are non-relativistic. In taking the non-relativistic limit, we have kept more terms in $p^0$ compared to $\omega_\mathbf{p}$ because we are here interested in the dispersion of the axion field which will give rise to qualitatively different effects. Also, we have performed a change of variables from the momentum distribution function $f(\mathbf{p})$ to the velocity distribution function $f(\mathbf{v})$, which in the non-relativistic limit is just a re-scaling by the axion rest mass $m_a$.

The velocity distribution $f(\mathbf{v})$ can in principle be any probability distribution, but a prominent model is the Standard Halo Model whereby $f(\mathbf{v})$ is given by the Maxwell-Boltzmann distribution in Eq. \eqref{eq:mb_velocity_dist}. For the sake of concreteness, we now assume that $f(\mathbf{v})$ is given by Eq.~\eqref{eq:mb_velocity_dist} and proceed to perform the angular integration in momentum space. To do so,  we choose a Cartesian coordinate system $\{\hat{s}, \hat{u}, \hat{v}\}$ such that $\hat{v}$ is parallel to $\mathbf{v}_E$ and $\hat{s}, \hat{u}$ are two other orthonormal basis vectors (see Fig.~\ref{fig:coord_schematic}). The $i$\textsuperscript{th} component of $\boldsymbol{\nabla}a$ is then
\begin{align}
\nonumber
\left(\boldsymbol{\nabla}a\right)^i(t) &= \sum_{jkl} \sqrt{\rho_\textrm{DM} f_{jk} v_j^2 \sin\theta_k \Delta \phi \Delta \theta \Delta v} \, \alpha_{jkl} \\
& \hspace{2cm} \times \cos\left(\omega_j t + \Phi_{jkl}\right) v^i,
\label{eq:axion_gradient_comp1}
\end{align}
where we have written the infinitesimal velocity volume $\Delta v^3$ in spherical coordinates and have discretized the velocity space into $N_j \times N_k \times N_l$ infinitesimal volumes at coordinates $(v_j, \theta_k, \phi_l)$. The frequency $\omega_j$ is defined as $\omega_j \equiv m_a(1+v_j^2/2)$ while $f_{jk} \equiv f(v_j, \theta_k, \phi_l) = f(v_j, \theta_k)$, and $v^i$ is related to $(v_j, \theta_k, \phi_l)$ by the standard transformations
\begin{equation}
v^i = 
\begin{cases}
v_j \sin\theta_k\cos\phi_l \quad&,\quad i=s \\
v_j \sin\theta_k \sin\phi_l \quad&,\quad i=u \\
v_j \cos\theta_k \quad&,\quad i=v
\end{cases}.
\end{equation}
Each infinitesimal volume, labeled by $jkl$, contributes for each $jkl$, an independent standard Rayleigh distributed random variable $\alpha_{jkl} \sim R(1)$, and an independent uniformly distributed random variable $\Phi_{jkl} \sim U(0, 2\pi)$ to the overall sum. We now re-write Eq. \eqref{eq:axion_gradient_comp1} as
\begin{align}
\nonumber
\left(\boldsymbol{\nabla}a\right)^i(t) &= \sum_{jk} \sqrt{\rho_\textrm{DM} f_{jk} v_j^2 \sin\theta_k \Delta \theta \Delta v}\, v_j W^i_k \\
& \hspace{1.5cm} \times (\cos\omega_j t \, C^i_{jk} - \sin\omega_j t \, S^i_{jk}),
\label{eq:axion_grad_azimuthal_out}
\end{align}
where
\begin{equation}
C^i_{jk} = 
\begin{cases}
\sqrt{\Delta\phi} \sum_l \alpha_{jkl} \cos \Phi_{jkl} \cos \phi_l \, &, \hspace{1mm} i=s \\
\sqrt{\Delta\phi} \sum_l \alpha_{jkl} \cos \Phi_{jkl} \sin \phi_l \, &, \hspace{1mm} i=u \\
\sqrt{\Delta\phi} \sum_l \alpha_{jkl} \cos \Phi_{jkl} \, &, \hspace{1mm} i=v \\
\end{cases},
\end{equation}
\begin{equation}
S^i_{jk} = 
\begin{cases}
\sqrt{\Delta\phi} \sum_l \alpha_{jkl} \sin \Phi_{jkl} \cos \phi_l \, &, \hspace{1mm} i=s \\
\sqrt{\Delta\phi} \sum_l \alpha_{jkl} \sin \Phi_{jkl} \sin \phi_l \, &, \hspace{1mm} i=u \\
\sqrt{\Delta\phi} \sum_l \alpha_{jkl} \sin \Phi_{jkl} \, &, \hspace{1mm} i=v \\
\end{cases},
\end{equation}
\begin{equation}
W^i_{k}=
\begin{cases}
\sin\theta_k \, &,\, i=s,u \\
\cos\theta_k \, &,\, i=v\\
\end{cases},
\end{equation}
and in the spirit of Monte-Carlo integration, we compute the azimuthal integral using the Central Limit Theorem, i.e. we let $\phi_l \to \phi_l \sim U(0, 2\pi)$, $N_l \to \infty$, and $\Delta\phi = 2\pi/N_l \to 0$. Since each $\alpha_{jkl}$, $\Phi_{jkl}$, and $\phi_l$ are identically and independently distributed, the summation over $l$ may, in the limit as $N_l \to \infty$, be evaluated using the Central Limit Theorem to give
\begin{equation}
C^i_{jk}=
\begin{cases}
\sqrt{\pi}\, x^i_{jk} \,&,\, i=s,u  \\
\sqrt{2\pi}\, x ^i_{jk} \,&,\, i=v \\
\end{cases},
\end{equation}
\begin{equation}
S^i_{jk}=
\begin{cases}
\sqrt{\pi}\, y^i_{jk} \,&,\, i=s,u  \\
\sqrt{2\pi}\, y ^i_{jk} \,&,\, i=v \\
\end{cases},
\end{equation}
where each $x^i_{jk}$ and $y^i_{jk}$ is an independent, standard normal-distributed variable
\begin{equation}
x^i_{jk} \sim N(0, 1) \quad,\quad y^i_{jk} \sim N(0, 1).
\end{equation}

It now remains to do the polar integral, which we perform by re-writing Eq. \eqref{eq:axion_grad_azimuthal_out} to become
\begin{equation}
\left(\boldsymbol{\nabla}a\right)^i(t) = \sum_j \sqrt{\rho_\textrm{DM} f_j \Delta v}\, v_j^2 (D^i_j \cos \omega_j t - T^i_j \sin \omega_j t),
\label{eq:axion_grad_polar_out}
\end{equation}
where we have separated out the polar dependence in $f_{jk}$ by defining
\begin{equation}
f_j \equiv \frac{1}{(2\pi \sigma_\textrm{v}^2)^{3/2}} \exp \left[-\frac{v_j^2 + v_E^2}{2\sigma_\textrm{v}^2}\right],
\label{eq:f_radial_discrete}
\end{equation}
and $D^i_j$, $T^i_j$ are given by
\begin{equation}
D^i_j = \sqrt{\Delta \theta} \sum_k \sqrt{\sin \theta_k e^{-\beta_j \cos \theta_k}} W^i_k C^i_{jk},
\end{equation}
\begin{equation}
T^i_j = \sqrt{\Delta \theta} \sum_k \sqrt{\sin \theta_k e^{-\beta_j \cos \theta_k}} W^i_k S^i_{jk},
\end{equation}
with $\beta_j$ defined as $\beta_j \equiv v_j v_E/\sigma_\textrm{v}^2$. The polar integral may be evaluated using the Central Limit Theorem as above, with $\Delta\theta=\pi/N_\theta$, $\theta_k \to \theta_k \sim U(0, \pi)$, and $N_\theta \to \infty$. By the Central Limit Theorem, $D^i_j$ and $T^i_j$ are then (independent) normally distributed random variables,
\begin{equation}
D^i_j,\, T^i_j \sim
\begin{cases}
N(0, 2\pi\psi_j) \, &, \, i=s,u \\
N(0, 4\pi\xi_j) \, &, \, i=v \\
\end{cases},
\end{equation}
where
\begin{equation}
\psi_j =
\begin{cases}
\dfrac{2(\beta_j \cosh \beta_j - \sinh \beta_j)}{\beta_j^3} \, &, \, \beta_j \neq 0 \\
\frac{2}{3} \, &, \, \beta_j = 0
\end{cases},
\end{equation}
\begin{equation}
\xi_j = 
\begin{cases}
\dfrac{(2 + \beta_j^2)\sinh \beta_j - 2 \beta_j \cosh \beta_j}{\beta_j^3} \, &, \, \beta_j \neq 0 \\
\frac{1}{3} \, &, \, \beta_j = 0
\end{cases}.
\end{equation}
Recognizing that $D^i_j$ and $T^i_j$ in Eq.~\eqref{eq:axion_grad_polar_out} are independent normally distributed variables with equal variances, we now re-write Eq. \eqref{eq:axion_grad_polar_out} in terms of a standard Rayleigh random variable $\alpha_{i, j}\sim R(1)$, and a uniformly distributed variable $\phi_{i, j}\sim U(0, 2\pi)$, to finally obtain an expression for the axion gradient without any angular dependence on the axion's velocity
\begin{align}
\nonumber
\left(\boldsymbol{\nabla}a\right)^i(t) &= \sum_j \sqrt{\rho_\textrm{DM} f_j \Delta v}\, v_j^2 \sqrt{\Var(D^i_j)} \alpha_{i, j} \\
\nonumber
& \hspace{2cm} \times\cos(\omega_j t + \phi_{i, j}) \\
&= \sum_j \sqrt{\pi \rho_\textrm{DM} f_j \Delta v}\, v_j^2 \epsilon_{i, j}\, \alpha_{i, j} \cos(\omega_j t + \phi_{i, j}),
\label{eq:axion_grad_radial}
\end{align}
where we have defined
\begin{equation}
\epsilon_{i, j} \equiv
\begin{cases}
\sqrt{2\psi_j} \quad ,\quad i=u,s \\
\sqrt{4\xi_j}\quad ,\quad i=v
\end{cases}.
\end{equation}

\subsection{Distribution of $A_k$ and $B_k$}

Armed with Eq. \eqref{eq:axion_grad_radial} for the axion's gradient that depends only on the magnitude of the axion's velocity, we are now ready to derive the stochastic properties of the experimental frequency spectrum, which as defined in Eq. \eqref{eq:Ak_Bk_def} and \eqref{eq:beta_fft_def}, is the discrete FFT of the anomalous field $\beta_n$ in Eq. \eqref{eq:beta_def}.

\begin{figure}
    \centering
    \includegraphics{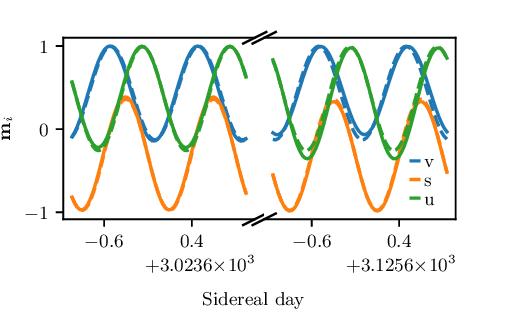}
    \caption{(color online) Components $s$ (orange), $u$ (green), and $v$ (blue) of the sensitive axis $\hat{\mathbf{m}}$ over the entire time span of the experiment. Solid lines give the exact results while dashed lines show fits of Eq. \eqref{eq:axis_approx} to the exact components.}
    \label{fig:sensitive_axis_fit}
\end{figure}

We begin by noting that to a good approximation, the time dependence of our experiment's sensitive axis $\hat{\mathbf{m}}$ over the $\sim$ 100 day span can be approximated by
\begin{equation}
\hat{\mathbf{m}}_i(n \Delta t)\approx C_i \cos(\omega_e n \Delta t +\theta_i) + D_i,
\label{eq:axis_approx}
\end{equation}
where $\omega_e$ is $2\pi$/(sidereal day) and $C_i, \theta_i$, and $D_i$ are obtained from fitting Eq.~\eqref{eq:axis_approx} to the actual $\hat{\mathbf{m}}_i(n \Delta t)$ during the span of the experiment, which can be calculated based on the orientation of the sensitive axis in the experiment (taken to be the outward normal to the surface of the Earth), the experiment's location on Earth (40.35 $^\circ$N, 74.65 $^\circ$W), and the absolute time that was measured as fractional sidereal days since J2000.0. Fig.~\ref{fig:sensitive_axis_fit} show the $\{s, u, v\}$ components of the sensitive axis, as well as fits of Eq.~\eqref{eq:axis_approx} to them, while Table~\ref{tab:sensitive_axis_comps} provides the values of $C_i$, $D_i$, and $\theta_i$ for $i=\{s,u,v\}$ obtained via the fits.

\begin{table}[!htb]
    \centering
    \begin{tabular*}{0.48\textwidth}{@{\extracolsep{\fill}} cccccc}
    \hline\hline
    $\boldsymbol{C_i}$ & \textbf{Value} & $\boldsymbol{\theta_i}$ & \textbf{Value} & $\boldsymbol{D_i}$ & \textbf{Value} \\
    \hline
    $C_s$ & 0.670 & $\theta_s$ & -1.550 & $D_s$ & -0.308 \\
    $C_u$ & 0.626 & $\theta_u$ & -2.739 & $D_u$ & 0.367 \\
    $C_v$ & 0.564 & $\theta_v$ & -0.492 & $D_v$ & 0.434 \\
    \hline\hline
    \end{tabular*}
    \caption{Fit parameters obtained via fitting Eq. \eqref{eq:axis_approx} to the exact components of the sensitive axis $\hat{\mathbf{m}}$.}
    \label{tab:sensitive_axis_comps}
\end{table}

If the data was collected with a sampling interval $\Delta t$ over a total interval $T \gg \Delta t$, such that the time series of the anomalous field $\beta_n$ ranges from $n=0,\ldots,N-1$, with $N\approx \Delta t/T$, then we may write
\begin{align}
\notag
\frac{2}{N}\beta_k &\approx 
\frac{2 \sqrt{\pi\rho_\textrm{DM}} g_\textrm{aNN}}{T \mu_\textrm{He}} \sum_{i=1}^3 \sum_{j=1}^M
\Delta t \sqrt{\Delta v {f}_{j}} {\alpha}_{i,j} {\epsilon}_{i,j} {v}_{j}^{2} \\
\nonumber
& \quad \times \sum_{n=0}^{N - 1} e^{- i \Delta t n \omega_{k}} \biggl[ \cos{\left(\Delta t n \omega_{j} + {\phi}_{i,j} \right)} {D}_{i} \\ 
\notag
& \quad\quad + \frac{\cos{\left(\Delta t n \omega_{e} - \Delta t n \omega_{j} - {\phi}_{i,j} + {\theta}_{i} \right)} {C}_{i}}{2} \\
& \left. \quad\quad + \frac{\cos{\left(\Delta t n \omega_{e} + \Delta t n \omega_{j} + {\phi}_{i,j} + {\theta}_{i} \right)} {C}_{i}}{2} \right] .
\label{eq:beta_k_1}
\end{align}
We note here that $\beta_k$ is the discrete Fourier transform of $\beta_n$ defined in Eq. \eqref{eq:beta_fft_def}, $\sum_i$ is over the three components of the $\{s, u, v\}$ basis while $\sum_j$ is over the speed of the axion and $\sum_n$ is over the time series. $M$, which is the number of slices we discretize the speed domain of the axion into, is arbitrary and we are therefore at liberty to take it as large as we like. The summation over $n$ may be evaluated using a geometric sum to give
\begin{align}
\nonumber
&\sum_{n=0}^{N-1} \Delta t \cos(\omega n \Delta t + \phi)e^{-i\omega_k n \Delta t} \\
&= 
\frac{\Delta t}{2}\left[\frac{1 - e^{i(\omega-\omega_k)\Delta t N}}{1 - e^{i(\omega-\omega_k)\Delta t}}e^{i\phi} + \frac{1 - e^{-i(\omega+\omega_k)\Delta t N}}{1 - e^{-i(\omega+\omega_k)\Delta t}}e^{-i\phi}
\right].
\end{align}
The terms in the bracket peak when $(\omega - \omega_k)\Delta t \approx 0$ or when $(\omega + \omega_k)\Delta t \approx 0$. For positive $\omega_k$ of interest, the first term dominates when $(\omega - \omega_k)\Delta t \approx 0$, and we may then write
\begin{align}
\nonumber
\sum_{n=0}^{N-1} \Delta t & \cos(\omega n \Delta t + \phi) e^{-i\omega_k n \Delta t} \\
&\hspace{2cm} \approx e^{i[(\omega - \omega_k)T/2 + \phi]}\frac{\sin((\omega-\omega_k)T/2)}{\omega-\omega_k}.
\end{align}
Practically, we would therefore require that our experimental sampling frequency $f_s \equiv 1/\Delta t$ is sufficiently high such that $2\pi \Delta f_a / f_s \ll 1$, where $\Delta f_a$ is the approximate full-width at half-maximum~(FWHM) of the axion peak. For $\Delta f_a \sim 10^{-6}$ Hz, this is easily achieved and the approximation is thus well satisfied. Applying this approximation, we have from Eq.~\eqref{eq:beta_k_1} and Eq.~\eqref{eq:Ak_Bk_def}, after expanding the trigonometric functions in Eq.~\eqref{eq:beta_k_1} and collecting the coefficients of $\cos(\phi_{i,j})$ and $\sin(\phi_{i,j})$,
\begin{align}
\nonumber
A_k &= -\frac{2 g_\textrm{aNN} \sqrt{\pi\rho_\textrm{DM}}}{T \mu_\textrm{He}} \sum_{i=1}^3\sum_{j=1}^M \sqrt{\Delta v f_j} \alpha_{i,j}\epsilon_{i,j}v_j^2 \\
& \hspace{1.5cm}\times \left[E_{i,j,k}\cos(\phi_{i,j}) + F_{i,j,k}\sin(\phi_{i,j})\right]
\label{eq:A_coeff_discrete}
\end{align}
\begin{align}
\nonumber
B_k &= -\frac{2 g_\textrm{aNN} \sqrt{\pi\rho_\textrm{DM}}}{T \mu_\textrm{He}} \sum_{i=1}^3\sum_{j=1}^M \sqrt{\Delta v f_j} \alpha_{i,j}\epsilon_{i,j}v_j^2 \\
& \hspace{1.5cm} \times \left[F_{i,j,k}\cos(\phi_{i,j}) - E_{i,j,k}\sin(\phi_{i,j})\right],
\label{eq:B_coeff_discrete}
\end{align}
where $E_{i,j,k}$ is given by
\begin{align}
\notag
& E_{i,j,k} = \\
\nonumber
& \frac{\sin{\left(\frac{T \omega_{e}}{2} + \frac{T \omega_{j}}{2} - \frac{T \omega_{k}}{2} \right)} \cos{\left(\frac{T \omega_{e}}{2} + \frac{T \omega_{j}}{2} - \frac{T \omega_{k}}{2} + {\theta}_{i} \right)} {C}_{i}}{2 \left(\omega_{e} + \omega_{j} - \omega_{k}\right)} \\
\notag
& + \frac{\sin{\left(\frac{T \omega_{j}}{2} - \frac{T \omega_{k}}{2} \right)} \cos{\left(\frac{T \omega_{j}}{2} - \frac{T \omega_{k}}{2} \right)} {D}_{i}}{\omega_{j} - \omega_{k}} - \\
& \frac{\sin{\left(\frac{T \omega_{e}}{2} - \frac{T \omega_{j}}{2} + \frac{T \omega_{k}}{2} \right)} \cos{\left(\frac{T \omega_{e}}{2} - \frac{T \omega_{j}}{2} + \frac{T \omega_{k}}{2} + {\theta}_{i} \right)} {C}_{i}}{2 \left(- \omega_{e} + \omega_{j} - \omega_{k}\right)},
\label{eq:E_ijk_def}
\end{align}
and $-F_{i,j,k}$ is
\begin{align}
\notag
&-F_{i,j,k} = \\
\nonumber
& \frac{\sin{\left(\frac{T \omega_{e}}{2} + \frac{T \omega_{j}}{2} - \frac{T \omega_{k}}{2} \right)} \sin{\left(\frac{T \omega_{e}}{2} + \frac{T \omega_{j}}{2} - \frac{T \omega_{k}}{2} + {\theta}_{i} \right)} {C}_{i}}{2 \left(\omega_{e} + \omega_{j} - \omega_{k}\right)} \\
\notag
& + \frac{\sin^{2}{\left(\frac{T \omega_{j}}{2} - \frac{T \omega_{k}}{2} \right)} {D}_{i}}{\omega_{j} - \omega_{k}} + \\
& \frac{\sin{\left(\frac{T \omega_{e}}{2} - \frac{T \omega_{j}}{2} + \frac{T \omega_{k}}{2} \right)} \sin{\left(\frac{T \omega_{e}}{2} - \frac{T \omega_{j}}{2} + \frac{T \omega_{k}}{2} + {\theta}_{i} \right)} {C}_{i}}{2 \left(- \omega_{e} + \omega_{j} - \omega_{k}\right)}.
\label{eq:F_ijk_def}
\end{align}
To see that $A_k$ and $B_k$ are normally distributed, we note that since $\alpha_{i,j}\sim R(1)$ and $\phi_{i,j}\sim U(0, 2\pi)$ are independent of each other for all $i, j$, we may re-write $\alpha_{i,j}\cos(\phi_{i,j}) \equiv x_{i,j} \sim N(0,1)$ and $\alpha_{i,j}\sin(\phi_{i,j}) \equiv y_{i,j} \sim N(0,1)$. Using this substitution, we obtain
\begin{align}
\nonumber
A_k &= -\frac{2 g_\textrm{aNN} \sqrt{\pi\rho_\textrm{DM}}}{T \mu_\textrm{He}} \sum_{i=1}^3\sum_{j=1}^M \sqrt{\Delta v f_j} \epsilon_{i,j}v_j^2 \\
& \hspace{3cm}\times \left[E_{i,j,k}\,x_{i,j} + F_{i,j,k}\,y_{i,j}\right]
\label{eq:A_coeff_discrete1}
\end{align}
\begin{align}
\nonumber
B_k &= -\frac{2 g_\textrm{aNN} \sqrt{\pi\rho_\textrm{DM}}}{T \mu_\textrm{He}} \sum_{i=1}^3\sum_{j=1}^M \sqrt{\Delta v f_j} \epsilon_{i,j}v_j^2 \\
& \hspace{3cm} \left[F_{i,j,k}\,x_{i,j} - E_{i,j,k}\,y_{i,j}\right].
\label{eq:B_coeff_discrete1}
\end{align}
We have thus shown that $A_k$ and $B_k$ are normally distributed random variables with zero mean. However, $A_k, B_k, A_r, B_r$ will in general have non-zero correlation with each other (note that $k$ and $r$ here index the fit frequency).
\subsection{Covariance of $A_k$ and $B_k$}
We therefore desire to compute the covariance matrix that will in general consist of
\begin{align}
\nonumber
\Cov(A_k, A_r) &= \Evalue(A_k A_r) \,,\, \Cov(A_k, B_r) = \Evalue(A_k B_r) \\
\Cov(B_k, A_r) &= \Evalue(B_k A_r) \,,\, \Cov(B_k, B_r) = \Evalue(B_k B_r).
\end{align}
It is important to note here that $x_{i, j}$ and $y_{i, j}$ in Eq.~\eqref{eq:A_coeff_discrete1} and Eq.~\eqref{eq:B_coeff_discrete1} are independent of each other for all $i, j$, while $x_{i,j}$ and $x_{p,q}$ are independent for all $i,j,p,q$ except when $i=p$ and $j=q$. Similarly, $y_{i,j}$ and $y_{p,q}$ are independent for all $i,j,p,q$ except when $i=p$ and $j=q$. This may be expressed succinctly as
\begin{align}
\nonumber
\Evalue(x_{i,j} x_{p,q}) = \delta_{ip}\delta_{jq} \,&,\, \Evalue(y_{i,j} y_{p,q}) = \delta_{ip}\delta_{jq} \\ \Evalue(x_{i,j} & y_{p,q}) = 0.
\end{align}
Using these identities, it is easy to show that the covariances are given by the following integrals
\begin{align}
\nonumber
\Cov(A_k, A_r) &= 4\pi\rho_\textrm{DM} \left(\frac{g_\textrm{aNN}}{T\mu_\textrm{He}}\right)^2 \sum_i \int_0^\infty dv \, f(v) \epsilon_i^2(v) \\
& \hspace{0.2cm} \times v^4\left[E_{ik}(v)E_{ir}(v) + F_{ik}(v)F_{ir}(v)\right]
\label{eq:covAA_def}
\end{align}
\begin{align}
\nonumber
\Cov(A_k, B_r) &= 4\pi\rho_\textrm{DM} \left(\frac{g_\textrm{aNN}}{T\mu_\textrm{He}}\right)^2 \sum_i \int_0^\infty dv \, f(v) \epsilon_i^2(v) \\
& \hspace{0.2cm} \times v^4 \left[E_{ik}(v)F_{ir}(v) - F_{ik}(v)E_{ir}(v)\right]
\label{eq:covAB_def}
\end{align}
\begin{align}
\nonumber
\Cov(B_k, A_r) &= - \Cov(A_k, B_r) = \Cov(A_r, B_k) \\
\Cov(B_k, B_r) &= \Cov(A_k, A_r),
\label{eq:covBB_def}
\end{align}
where we have taken the continuum limit by letting $M\to\infty$ and thus $\sum\Delta v \to \int dv$ and $v_j \to v$, which implies that $\omega_j \equiv m_a + m_a v_j^2/2 \to \omega(v) \equiv m_a + m_a v^2/2$ and $E_{ijk} \to E_{ik}(v)$, $F_{ijk} \to F_{ik}(v)$, $f_j \to f(v)$, $\epsilon_{i,j}\to\epsilon_i(v)$. Although closed form expressions for Eq.~\eqref{eq:covAA_def}, \eqref{eq:covAB_def}, and \eqref{eq:covBB_def} may not exist, they can always be evaluated numerically.

\subsection{Infinite frequency resolution limit}

In the limit that $T \to \infty$, the integrands in Eq.~\eqref{eq:covAA_def} and \eqref{eq:covAB_def} become increasingly oscillatory and difficult to evaluate numerically, but thankfully, closed form solutions exist in that limit. To obtain those solutions, we note that in the continuum limit and as $T \to \infty$,
\begin{align}
\nonumber
E_{ik}(v) &\approx \frac{\pi}{m_a}\left[
\frac{C_i \cos\theta_i}{2\aspeed{k}{-1}}\delta (v - \aspeed{k}{-1}) + \right.\\
&\left.\hspace{1.2cm}
\frac{C_i \cos\theta_i}{2\aspeed{k}{1}} \delta (v - \aspeed{k}{1}) +
\frac{D_i}{\aspeed{k}{0}} \delta (v - \aspeed{k}{0})
\right] \label{eq:E_ik_limit} \\
F_{ik}(v) &\approx \frac{\pi C_i \sin \theta_i}{2 m_a}\left[
\frac{\delta(v - \aspeed{k}{1})}{\aspeed{k}{1}} - \frac{\delta(v - \aspeed{k}{-1})}{\aspeed{k}{-1}}
\right] \label{eq:F_ik_limit},
\end{align}
since $\lim_{\epsilon\to0}\, \sin(z/\epsilon)/z = \pi \delta(z)$. Physically, $\aspeed{k}{n}$ is the speed of an axion with mass $m_a$ oscillating at a frequency $\omega_k$, as measured from the $n$\textsuperscript{th} sideband with $n\in\{-1,0,1\}$, and is given by Eq.~\eqref{eq:axion_speed}. As explained in Sec.~\ref{sec:freq_domain}, these sidebands originate from sidereal modulation of the experiment's sensitive axis as the Earth rotates about its axis.

Now we would like to make use of the Dirac delta functions in Eq.~\eqref{eq:E_ik_limit} and \eqref{eq:F_ik_limit} to obtain an analytical form of the covariance matrix. However, the integrals in Eq.~\eqref{eq:covAA_def} and \eqref{eq:covAB_def} contain factors like $E_{ik}E_{ir}$ and $E_{ik}F_{ir}$, which means that we would end up with ill-defined terms containing two Dirac delta functions. To avoid this, we first take the continuum limit and integrate over $v$ in Eq.~\eqref{eq:A_coeff_discrete1} and \eqref{eq:B_coeff_discrete1} before computing the covariance matrix elements from $A_k$ and $B_k$. Although this approach avoids having to integrate over a product of Dirac delta functions, one must take the continuum limit carefully such that there is an integration measure of $\Delta v$ and not $\sqrt{\Delta v}$. One way to do this is to note that a velocity grid $\{v_j\}$ has a corresponding (angular) frequency grid $\{\omega_j \, | \, \omega_j = m_a + m_a v_j^2/2\}$. The spacing of this frequency grid is $\Delta \omega \approx m_a v_j \Delta v$. If we choose this spacing to be $\Delta \omega = 2\pi/T$, then we have
\begin{equation}
\frac{1}{\sqrt{T}} \approx \sqrt{\frac{m_a v_j}{2\pi}\Delta v}.
\end{equation}
Consequently, we may write the continuum limit of Eq.~\eqref{eq:A_coeff_discrete1} and \eqref{eq:B_coeff_discrete1} as
\begin{align}
\nonumber
A_k &= -\frac{g_\textrm{aNN}}{\mu_\textrm{He}} \sqrt{\frac{2m_a\rho_\textrm{DM}}{T}} \sum_{i=1}^3 \int_0^\infty\,dv \sqrt{f(v)}\epsilon_i(v) v^{5/2} \\
& \hspace{2cm} \times \left[E_{ik}(v)x_i(v) + F_{ik}(v)y_i(v)\right]
\end{align}
\begin{align}
\nonumber
B_k &= -\frac{g_\textrm{aNN}}{\mu_\textrm{He}} \sqrt{\frac{2m_a\rho_\textrm{DM}}{T}} \sum_{i=1}^3 \int_0^\infty\,dv \sqrt{f(v)}\epsilon_i(v) v^{5/2} \\
& \hspace{2cm} \times \left[F_{ik}(v)x_i(v) - E_{ik}(v)y_i(v)\right].
\end{align}
Substituting in the approximations Eq. \eqref{eq:E_ik_limit} and \eqref{eq:F_ik_limit}, we obtain after integrating over $v$
\begin{align}
\nonumber
&A_k \approx -\frac{\pi g_\textrm{aNN}}{m_a\mu_\textrm{He}}\sqrt{\frac{2m_a\rho_\textrm{DM}}{T}} \sum_{i=1}^3\\ 
\nonumber
& \left[
\frac{C_i\cos\theta_i}{2} g_i(\aspeed{k}{-1}) x(\aspeed{k}{-1}) + \frac{C_i\cos\theta_i}{2} g_i(\aspeed{k}{1}) x(\aspeed{k}{1}) \right. \\
\nonumber
& \, + D_i \, g_i(\aspeed{k}{0}) x(\aspeed{k}{0}) + \frac{C_i\sin\theta_i}{2} g_i(\aspeed{k}{1}) y(\aspeed{k}{1}) \\
& \left. \, - \frac{C_i\sin\theta_i}{2} g_i(\aspeed{k}{-1}) y(\aspeed{k}{-1}) \right]
\end{align}
\begin{align}
\nonumber
&B_k \approx -\frac{\pi g_\textrm{aNN}}{m_a\mu_\textrm{He}}\sqrt{\frac{2m_a\rho_\textrm{DM}}{T}}\sum_{i=1}^3 \\
\nonumber
& \left[
-\frac{C_i\cos\theta_i}{2} g_i(\aspeed{k}{-1}) y(\aspeed{k}{-1}) - \frac{C_i\cos\theta_i}{2} g_i(\aspeed{k}{1}) y(\aspeed{k}{1}) \right. \\
\nonumber
& \, -D_i g_i(\aspeed{k}{0}) y(\aspeed{k}{0}) + \frac{C_i\sin\theta_i}{2} g_i(\aspeed{k}{1}) x(\aspeed{k}{1}) \\
& \left. \, - \frac{C_i\sin\theta_i}{2} g_i(\aspeed{k}{-1}) x(\aspeed{k}{-1}) \right],
\end{align}
where we have for brevity defined
\begin{equation}
g_i(v) \equiv \sqrt{f(v)} \epsilon_i(v) v^{3/2}.
\end{equation}
The covariance matrix may now be computed by remembering that $x(\nu)$ and $y(\nu)$ are independent standard normal variables for all $\nu$, while $x(\nu_1)$ is independent from $x(\nu_2)$ for all $\nu_1\neq\nu_2$ and $y(\nu_1)$ is independent from $y(\nu_2)$ for all $\nu_1\neq\nu_2$. More precisely, we have
\begin{align}
\nonumber
\Cov(x(\nu_1)x(\nu_2)) = \delta_{\nu_1,\nu_2} \,&,\, \Cov(y(\nu_1)y(\nu_2)) = \delta_{\nu_1,\nu_2} \\ \Cov(x(\nu_1)& y(\nu_2))=0.
\end{align}
Using these properties, the variance of $A_k$ and $B_k$ can be computed to give
\begin{align}
\nonumber
&\Var(A_k)=\Var(B_k)=\left(\frac{g_\textrm{aNN}}{\mu_\textrm{He}}\right)^2 \frac{\pi\rho_\textrm{DM}\,\Delta f}{f_a} \\
&\, \times\sum_{i=1}^3
\left[\frac{C_i^2}{4} g_i^2(\aspeed{k}{-1}) + D_i^2 g_i^2(\aspeed{k}{0}) + \frac{C_i^2}{4} g_i^2(\aspeed{k}{1})\right],
\end{align}
where we have $\Delta f \equiv 1/T$ and $f_a=m_a/(2\pi)$ is the axion mass frequency. Similarly, the covariance $\Cov(A_kA_r)$ and $\Cov(B_kB_r)$, for $k\neq r$ is
\begin{align}
\nonumber
&\Cov(A_kA_r)= \left(\frac{g_\textrm{aNN}}{\mu_\textrm{He}}\right)^2 \frac{\pi\rho_\textrm{DM}\,\Delta f}{f_a}
\sum_{i=1}^3 \\
\nonumber
&\left( \chi_{i,k,-1}[\aspeed{k}{-1}=\aspeed{r}{1}] + \sum_{m=-1}^0 \eta_{i,k,m}[\aspeed{k}{m}=\aspeed{r}{m+1}] \right. \\
\nonumber
& \left.\quad + \chi_{i,k,1} [\aspeed{k}{1}=\aspeed{r}{-1}] + \sum_{m=0}^1 \eta_{i,k,m} [\aspeed{k}{m}=\aspeed{r}{m-1}]\right) \\
&=\Cov(B_kB_r),
\end{align}
and $\Cov(A_kB_r)$ for all $k,r$ is
\begin{align}
\nonumber
&\Cov(A_kB_r)= \left(\frac{g_\textrm{aNN}}{\mu_\textrm{He}}\right)^2 \frac{\pi\rho_\textrm{DM}\,\Delta f}{f_a}
\sum_{i=1}^3 \\
\nonumber
&\left( \zeta_{i,k,-1} [\aspeed{k}{-1}=\aspeed{r}{1}] + \sum_{m=-1}^0 \kappa_{i,k,m}[\aspeed{k}{m}=\aspeed{r}{m+1}] \right. \\
& \left. \quad - \zeta_{i,k,1}[\aspeed{k}{1}=\aspeed{r}{-1}] -\sum_{m=0}^1 \kappa_{i,k,m}[\aspeed{k}{m}=\aspeed{r}{m-1}] \right),
\end{align}
where the square brackets $[\ldots]$ above denote the Iverson bracket, and we have defined
\begin{align}
\chi_{i,k,m} &\equiv\frac{C_i^2\cos2\theta_i}{4} g_i^2(\aspeed{k}{m}) \\
\eta_{i,k,m} &\equiv\frac{C_i D_i \cos\theta_i}{2} g_i^2(\aspeed{k}{m}) \\
\zeta_{i,k,m} &\equiv\frac{C_i^2 \sin2\theta_i}{4} g_i^2(\aspeed{k}{m}) \\
\kappa_{i,k,m} &\equiv\frac{C_i D_i \sin\theta_i}{2} g_i^2(\aspeed{k}{m}).
\end{align}

It is frequently useful to know the expected power spectrum of the axion, which in the continuum limit is defined as $\Evalue[R^2] = \Evalue[A^2(\omega) + B^2(\omega)]$. In the limit of infinite frequency resolution, this can be, using similar techniques as above, evaluated to give
\begin{align}
\nonumber
\Evalue[R^2(\omega)] &= \frac{2\pi\rho_\textrm{DM}\,\Delta f \,g_\textrm{aNN}^2}{ f_a \mu_\textrm{He}^2} \sum_{i=1}^{3} \\
\nonumber
&\quad\left(\dfrac{{C}_{i}^{2} \epsilon_i^2(\nu_{-1}(\omega)) \tilde{f}(\nu_{-1}(\omega)) \nu_{-1}^3(\omega)}{4} \right. \\
\nonumber
& \quad\quad + {D}_{i}^{2} \epsilon_i^2(\nu_0(\omega)) \tilde{f}(\nu_0(\omega)) \nu^3_0(\omega) \\
&\quad\quad + \left. \dfrac{{C}_{i}^{2} \epsilon_i^2(\nu_1(\omega)) \tilde{f}(\nu_1(\omega)) \nu_1^3(\omega)}{4} \right),
\label{eq:R2_avg}
\end{align}
where
\begin{equation}
    \nu_{n}(\omega) = \begin{cases}
    \sqrt{\dfrac{2 (\omega - (m_a + n \omega_e))}{m_a}} &, \omega - (m_a + n \omega_e) \geq 0 \\
    0 &, \omega - (m_a + n \omega_e) < 0
    \end{cases},
    \label{eq:axion_speed_continuum}
\end{equation}
for $n=-1,0,1$ is the continuum version of Eq. \eqref{eq:axion_speed}, and
\begin{equation}
\tilde{f}(v) \equiv \frac{1}{(2\pi \sigma_\textrm{v}^2)^{3/2}} \exp \left[-\frac{v^2 + v_E^2}{2\sigma_\textrm{v}^2}\right],
\label{eq:f_radial_continuum}
\end{equation}
is the continuum version of Eq. \eqref{eq:f_radial_discrete}.

\section{Frequency grid spacing\label{sec:freq_grid}}

We claimed in Sec.~\ref{sec:ll_procedure} that it is necessary in our formalism to test for axions with a frequency grid of spacing $\approx \Delta f_a/2$ to correctly recover or set upper limits on $g_\textrm{aNN}$ from an axion signal within our mass range of interest. We now justify this claim below.

\begin{figure}[t]
    \centering
    \includegraphics[width=0.49\textwidth]{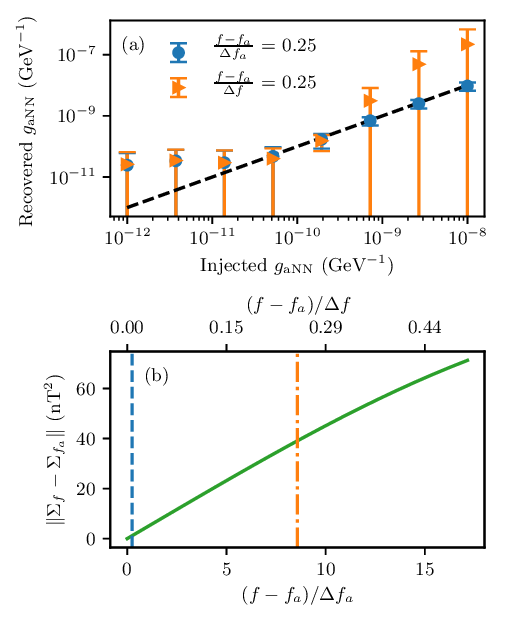}
    \caption{(color online) (a) Signal injection and recovery plots for the case when the test frequency $f$ does not equal the true axion frequency $f_a$. The blue marker shows the case when $f$ is a quarter of the axion linewidth $\Delta f_a$ away from the true axion frequency, while the orange marker shows the case when $f$ is misaligned by a quarter of the frequency grid spacing $\Delta f$. We present here for $\Delta f_a \ll \Delta f$. (b) Frobenius norm of the difference between the test covariance matrix $\Sigma_f$ and the true axion covariance matrix $\Sigma_{f_a}$. The blue dashed vertical line corresponds to the legend of the blue marker in (a) while the orange dash-dot vertical line corresponds to the legend of the orange marker in (a).}
    \label{fig:freq_point_shift}
\end{figure}

In Fig.~\ref{fig:freq_point_shift}a, we show the recovered $g_\textrm{aNN}$ from Monte-Carlo datasets with injected axion signals of frequency $f_a$ and varying $g_\textrm{aNN}$. At each value of injected $g_\textrm{aNN}$, we test for an axion with mass $f \neq f_a$ and attempt to recover $g_\textrm{aNN}$ from 100 Monte-Carlo datasets. Markers denote the average $g_\textrm{aNN}$ recovered from all 100 Monte-Carlo datasets while vertical lines give the standard deviation of the recovered $g_\textrm{aNN}$. We show the results when the test frequency $f$ is displaced from the actual axion frequency $f_a$ by $\Delta f_a/4$~(blue) and $\Delta f/4$~(orange).

As the large standard deviation of the recovered best-fit values of $g_\textrm{aNN}$ (orange triangle markers) in Fig.~\ref{fig:freq_point_shift}a show, testing at the scale of the experimental resolution $\Delta f$ can lead to wrong results even when the test frequency $f$ is separated from the true axion frequency $f_a$ by only a quarter of $\Delta f$ if the axion linewidth $\Delta f_a$ is sufficiently narrow compared to the frequency grid spacing ($\Delta f/\Delta f_a \approx 34$ in Fig.~\ref{fig:freq_point_shift}).
This is because given a fixed frequency grid with spacing defined by the experimental frequency resolution, the expected power (and correlation) spectrum on that grid from a true axion signal changes smoothly as a function of the axion mass. Consequently, if there is sufficient discrepancy between the measured power (and correlation) spectrum coming from a real axion at frequency $f_a$ compared to the expected power (and correlation) spectrum coming from an axion at test frequency $f\neq f_a$, the likelihood analysis will correctly conclude that there is no axion at test frequency $f$. One way to visualize this is to plot, as in Fig.~\ref{fig:freq_point_shift}b, the Frobenius norm of the difference between the correlation matrix $\Sigma_f$ of an axion at test frequency $f$ and the correlation matrix $\Sigma_{f_a}$ from the actual axion at frequency $f_a$ (with both matrices defined on the same frequency grid and using the same value of $g_\textrm{aNN}$). As the orange dash-dot line in Fig.~\ref{fig:freq_point_shift}b shows, the test frequency $f$ differs from $f_a$ by only $\Delta f/4$ (upper $x$-axis), but it differs from $f_a$ by $\approx 8 \Delta f_a$ (bottom $x$-axis), and the Frobenius norm of the difference in the correlation matrices is about 40 nT$^2$, which is sufficient for the likelihood analysis to conclude that there is no axion at test frequency $f$ as seen by the diverging orange standard deviations of the recovered best-fit $g_\textrm{aNN}$ in Fig.~\ref{fig:freq_point_shift}a.

On the other hand, for a sufficiently small separation between $f$ and $f_a$ such that the difference between their respective correlation matrices is nearly zero, we would expect that the likelihood analysis will be unable to differentiate between the two and would therefore recover $g_\textrm{aNN}$ from an axion at frequency $f_a$ as though it were at test frequency $f$. This is demonstrated by the blue circular markers in Fig.~\ref{fig:freq_point_shift}a, which shows the likelihood analysis recovering the injected $g_\textrm{aNN}$ from an actual axion at frequency $f_a$ while testing at frequency $f$ when $f - f_a = \Delta f_a/4$ is sufficiently small such that the Frobenius norm of the difference in their correlation matrices is almost zero (see blue dashed line in Fig. \ref{fig:freq_point_shift}b). To put it another way, axion masses need to be tested at intervals of around $\Delta f_a/2$ so that if there \textit{were} an axion signal, the furthest test frequency would be about $\Delta f_a/4$ away and the analysis would correctly recover $g_\textrm{aNN}$ from a real signal. Monte-Carlo simulations across the frequency range of our analysis from 0.01 to 10 Hz indicates that a spacing of $\Delta f_a/2$ is adequate in that it recovers, within one standard deviation (taken over the ensemble of Monte-Carlo simulations), the correct injected $g_\textrm{aNN}$ value.


\providecommand{\noopsort}[1]{}\providecommand{\singleletter}[1]{#1}%

\end{document}